分类号______________  密级______________

ＵＤＣ______________  编号______________

# 清 华 大 学

# 博 士 后 研 究 工 作 报 告

半导体-金属复合圆盘腔纳米激光器
________________________________________

________________________________________

张泰平

工作完成日期　　2016 年 1 月—2018 年 3 月
　　　　　　　　________________________

报告提交日期　　　　2018 年 3 月
　　　　　　　　________________________

清 华 大 学 （北京）

2018 年 3 月

# 半导体-金属复合圆盘腔纳米激光器

# SEMICODUCTOR-METALLIC HYBRID NANO-DISK LASER

博　士　后　姓　名　张泰平

流动站（一级学科）名称　清华大学电子科学与技术

专　业（二级学科）名称　物理电子学

研究工作起始时间　2016 年 1 月 14 日

研究工作期满时间　2018 年 4 月 14 日

# 摘 要

本研究设计、制备了基于 InGaAsP 多量子阱和 InGaAs 体材料薄膜的半导体-金属复合圆盘腔纳米激光器，并且对激光器的激光性能进行了表征。通过在半导体-金属复合腔纳米激光器的设计中在增益介质与金属壳层之间加入一薄层透明介质层有效减低金属损耗，提高了表面等离极化激元模式的品质因子。InGaAsP 多量子阱半导体-金属复合圆盘腔纳米激光器在室温下可以产生表面等离极化激元模式的激光。器件的直径约为 600 nm，激光模式为类 TM 模表面等离极化激元模式。在高泵浦功率下，纳米激光器的多量子阱势阱能级被填充饱和，载流子会填充势阱的子带能级和势垒能级，可以先后产生多个波长的激光。以 InGaAs 体材料半导体-金属复合圆盘腔纳米激光器在室温下同样存在低能级填充饱和，高能级被填充的现象，可以产生多个波长的激射。而在 77 K 的低温光学表征中，直径 400 nm 的器件可以产生类 TE 模表面等离极化激元模式的激光。

关键词：纳米激光器，半导体-金属复合微腔，表面等离极化激元模式，多模激光

# Abstract

In this research, semiconductor-metallic hybrid nano-disk laser based on InGaAsP multi quantum wells (MQWs) or InGaAs bulk material membranes were designed, fabricated and characterized. The quality factors of surface-plasmon-polariton (SPP) modes can be improved by introducing a transparent dielectric shield layer between the metallic cap and the gain material. The InGaAsP MQWs based semiconductor-metallic hybrid nano-laser can surpport SPP mode lasing at room temperature. With diameter of 600 nm, the lasing mode is TM-like SPP mode. Under high pumping, for nano-laser, the quantized levels in the wells and the barrier are populated. This behavior supported multi-mode lasing. The InGaAs bulk material based semiconductor-metallic hybrid nano-laser perform that the carrier density saturated the energy band and populated the higher band as well. The nano-laser also supports multi-mode lasing at room temperature. At loe temperature of 77K, the nano-laser with 400 nm diameter can support TE-like SPP mode lasing.

Key words: Nano-laser, semiconductor-metallic hybrid micro cavity, surface-plasmon-polariton (SPP) modes, multi-mode lasing

# 目 录

# 第一章 绪 论

## 1.1 研究背景

自 1960 年第一个激光器诞生以来,就对人类的生活和科技的发展产生了重大的影响。其中,半导体激光器在光电子技术,信息技术等领域已得到广泛应用。随着集成光电子技术的发展,光子学器件相对电子器件而言较大的器件尺寸成为了阻碍高密度集成的关键问题之一。而作为集成光子学芯片中光源的理想候选者,半导体激光器的器件小型化也成为近年来被广泛关注的研究方向[1]。基于纳米材料制备技术和微纳加工技术的迅速发展,多种类型的纳米激光器在实验研究中得以实现 [1]。纳米激光器具有低功耗,高调制速度等潜在优势,在数据存储、生物传感、片上光通信和光计算等领域具有潜在应用价值 [2]。对于纳米激光器而言,随着器件尺寸的减小,光的波长成为器件尺寸进一步减小的主要限制因素[1]。这一因素的存在使得具有单纯半导体谐振腔的激光器在尺寸接近或小于所涉及波长时,谐振腔对光的限制能力快速减弱。金属材料的反射作用以及表面等离极化激元效应对光场具有良好的限制能力,因此在谐振腔结构中引入金属结构成为提高谐振腔对光的限制能力,进一步减小器件尺寸的有效策略。在半导体激光器小型化的历程中,经历了由半导体或电介质谐振腔到半导体-金属复合谐振腔的结构转变,就是金属材料优秀的光场限制能力被应用的实例。至今,半导体-金属复合腔纳米激光器已经实现了电注入激射。但是,电注入器件的尺寸还没有实现在三个维度上同时小于所涉及的波长。研究工作者依然在不懈的致力于器件的小型化研究,以实现在三维尺度上达到亚波长甚至小于半波长的激光器。同时,对于纳米激光器表现出的新颖的器件物理性能的研究工作也在逐渐展开。

## 1.2 纳米激光器的研究进展

在半导体-金属复合腔纳米激光器被报道之前,纳米激光器的研究主要集中于半导体微腔激光器,如纳米线激光器,微盘激光器和光子晶体微腔激光器等。其中,纳米线激光器由于其新颖的材料体系和极小的器件体积,吸引了研究者的注意。导体纳米线的材料如 ZnO,CdS 等具有可形成粒子数反转的性能,两端相互平行的截面则形成了天然的 F-P 光学谐振腔的结构,而且纳米线和空气很大的折射率差也提高了。这样就使这些纳米线具备了可以被制备成激光器件的条件。对于单纯纳米线激光器的研究中,杨培东研究组利用蓝宝石外延衬底成功合成 ZnO 纳米线阵列,实现了低阈值室温紫外激光发射 [3]。之后该研究组又通过用激光照射单根 ZnO 纳米线,从而获得了单根 ZnO 纳米线的光致激光器件的激光发射 [4]。该小组又使用探针将平置于 Si 衬底上的 GaN 纳米线弯成环形，从而得到了 GaN 纳米线环形谐振腔，并发现相对于伸展的纳米线样品发射的激光，纳米线环形谐振腔发射的激光波长发生了红移，并且随着环形直径的减小，红移的幅度增加 [5]。C. M. Lieber 小组通过用激光照射单根 CdS 纳米线获得了单根纳米线的纳米线光致激光器件 [6]。C. Ronning 小组制备了锡掺杂的 CdS 纳米线，得到了可由连续激光激发的纳米线光致激光器件 [7]。在将金属结构引入激光器设计后，与纳米线激光器相关的研究工作实现了具有更小模式体积和器件体积的表面等离极化激元模式的激光发射，对其他结构纳米激光器的表面等离极化激元模式激光的实现提供了启发。该领域的代表性工作有张翔研究组将 CdS 纳米线平置于沉积有一层 $MgF_2$ 作为隔离层的 Au 薄膜上。通过纳米线激光与 Au 膜的表面等离极化激元的耦合，获得了表面等离极化激元激光发射 [8]。这一研究对利用表面等离极化激元产生激光具有重要的启发意义。T. C. Sum 研究组以 $SiO_2$ 作为隔离层，同样制备出了 CdS 纳米线激光与 Au 膜表面等离极化激元激光发射，并发现随着隔离层厚度的减小，表面等离极化激元增强了 Burstein−Moss 效应，引发了激光波长的明显蓝移，从而为激光波长的调节提供了可行的途径 [9]。

纳米线激光器的激光由于材料种类限制，只能产生紫外线至可见光波长范围的激光。对于通信常用的 1300 nm 或 1500 nm 附近波长的激光而言，较有效的增益材料依然是基于 III-V 族材料体系。因为该体系材料尚未有类似 ZnO、CdS 等材料的简易的生长纳米线的方法，所以基于 III-V 族材料的半导体激光器依然是使用微盘腔，光子晶体微腔等结构 [10-16]。其中光子晶体腔结构的器件尺寸难以实现亚微米化，因此，基于 III-V 族增益材料的半导体激光器微型化研究集中于具有微盘腔、微柱腔等结构的器件。在这一进程中，引入金属结构的益处展现的尤为明显。理论上，我们可以通过图 1-1 所示的文献 [17] 中的数值仿真对金属结构对微盘腔纳米器件的光学性能的影响得到一个直观的结论。图中，左图展示了单纯半导体圆盘腔和侧壁包覆一层银薄膜的半导体-金属复合圆盘腔对应于 1550 nm 波长光的模式随着圆盘半径的减小品质因子的变化。可以发现单纯半导体圆盘腔的品质因子随其半径减小迅速下降，当圆盘半径从 700 nm 减小到 350 nm，模式的品质因子从 52884 减小到了 92。而包覆银薄膜的圆盘腔在圆盘半径为 340 nm 时，品质因子仍可达到 722。甚至在圆盘半径减小到 150 nm 时，其品质因子仍然达到 326。这一结果表明在小尺寸下覆有金属薄膜的圆盘腔器件具有更高的品质因子。而右图通过两种微腔的品质因子的变化趋势进行比较，更直观的反映出在微腔直径

减小的过程中单纯半导体腔由于对光的束缚能力减小导致微腔品质因子急速下降,而包覆银薄膜的微腔则依靠金属对光优秀的限制能力非常有效的减缓了品质因子减小的速度。这一理论依据使通过在器件设计中引入金属结构实现半导体激光器的进一步微型化具有了可行性。

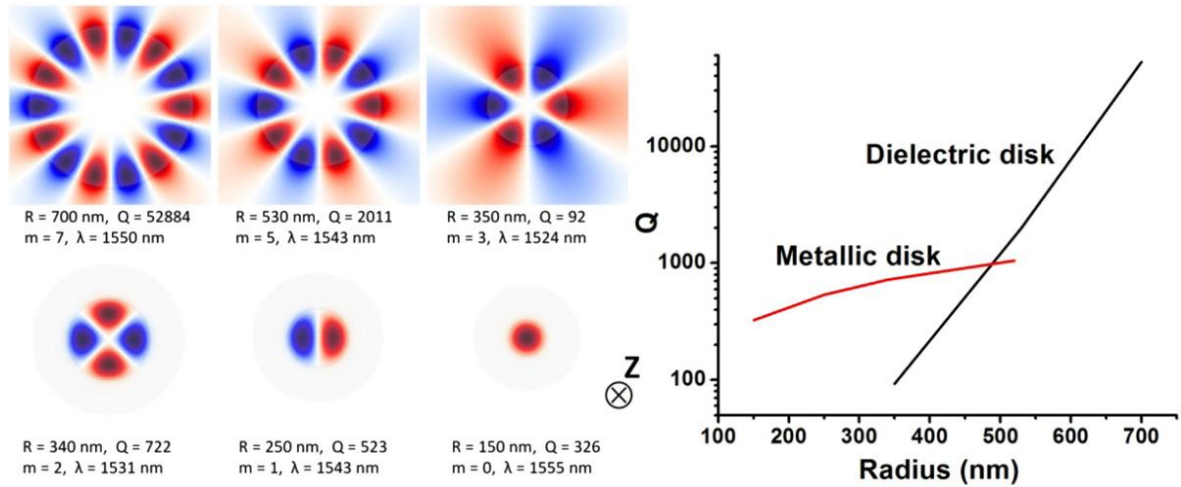

图 1-1 单纯半导体圆盘腔直径和包覆银薄膜的半导体圆盘腔在 1500 nm 波长附近的模式的品质因子随圆盘半径减小的数值变化 (左)与两种微腔的模式的品质因子与圆盘半径的关系曲线 (右) [17]。

2007 年,M. T. Hill 等人报道了第一个半导体-金属复合腔纳米激光器,该器件采用如图 1-2 所示的圆柱腔结构,在 77 K 的温度下通过电注入的方式得到激射 [18]。此后报道的电注入的纳米激光器均采用了由此结构演化的 MISIM (metal-insulator-semiconductor-insulator-metal)结构。此结构中,绝缘层在起到绝缘作用之外,还起到了隔离金属层与有源区,提高品质因子的作用。在电注入器件方向,宁存政课题组长期致力于器件微型化的研究。至 2013 年,报道了如图 1-3 所示的基于矩形截面 MISIM 结构柱形腔激光器,实现了室温下线宽仅 0.5 nm 的电注入连续波激射 [19]。

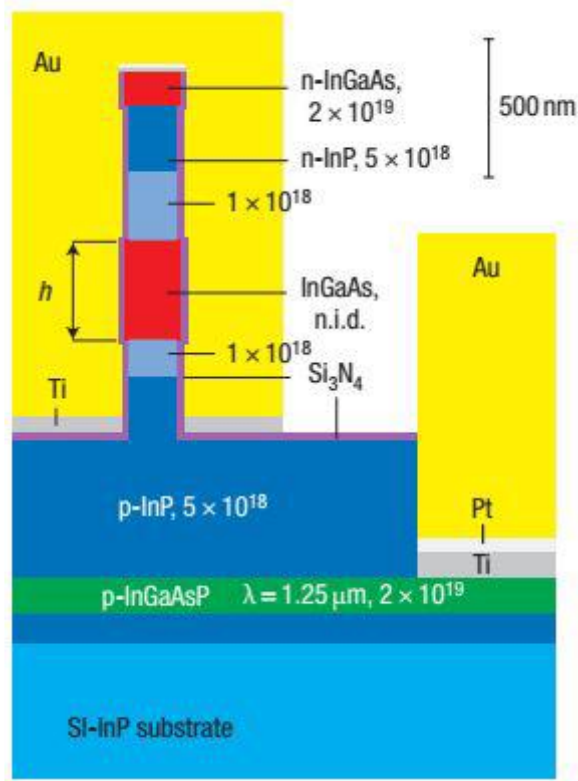

图 1-2 半导体-金属圆柱腔纳米激光器结构 [18]。

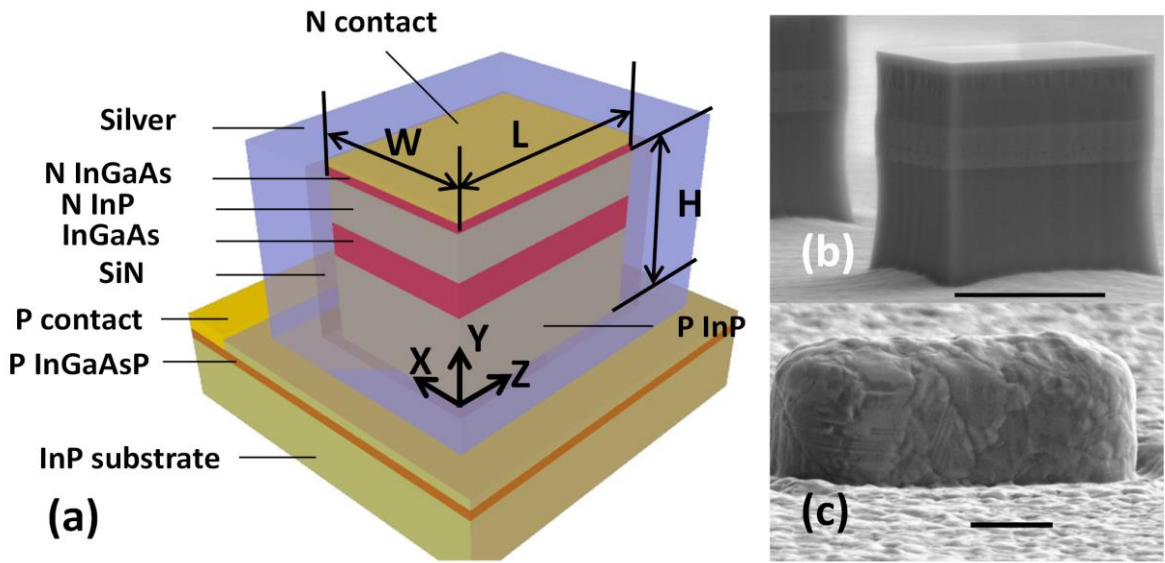

图 1-3 矩形截面 MISIM 柱形腔激光器 [19]。

基于厚度在 1 μm 以下的薄膜材料的半导体-金属复合腔纳米激光器一般具有圆盘或圆环腔结构，至今仍然以光泵浦激发为主。具有代表性的报道有 Y. Fainman 课题组开发了多量子阱圆环腔结构的半导体-金属纳米激光器，得到了圆环直径 500 nm 的器件的室温和低温激射 [20]。M. C.

Wu 课题组开发了基于体材料的三明治结构的半导体-金属纳米激光器，实现了圆盘直径 400 nm 器件的 77 K 温度下的激射 [21]。H-G Park 课题组开发了基于多量子阱圆盘腔结构的半导体-金属纳米激光器，器件结构如图 1-4 所示，金属壳层紧密包覆增益介质圆盘。报道分析了该结构器件所具有如图 1-5 所示的类 TM 模和类 TE 模两种表面等离极化激元模式和介质光学模式。该研究在实验中实现了直径 1 μm 器件在 9 K~77 K 温度下的类 TM 模表面等离极化激元模式的脉冲激射 [22,23]。

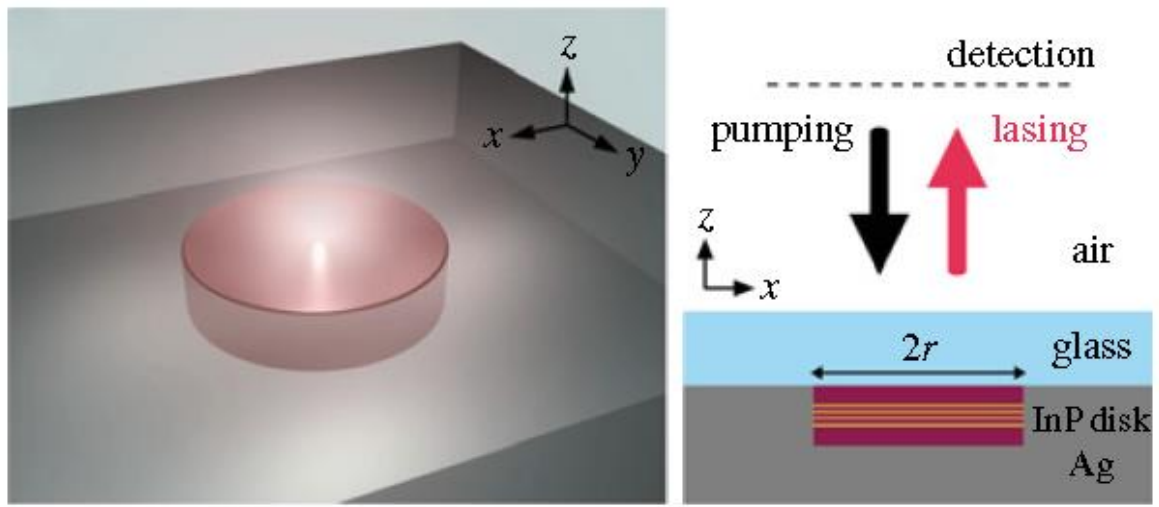

图 1-4 多量子阱圆盘腔半导体-金属纳米激光器 [22,23]。

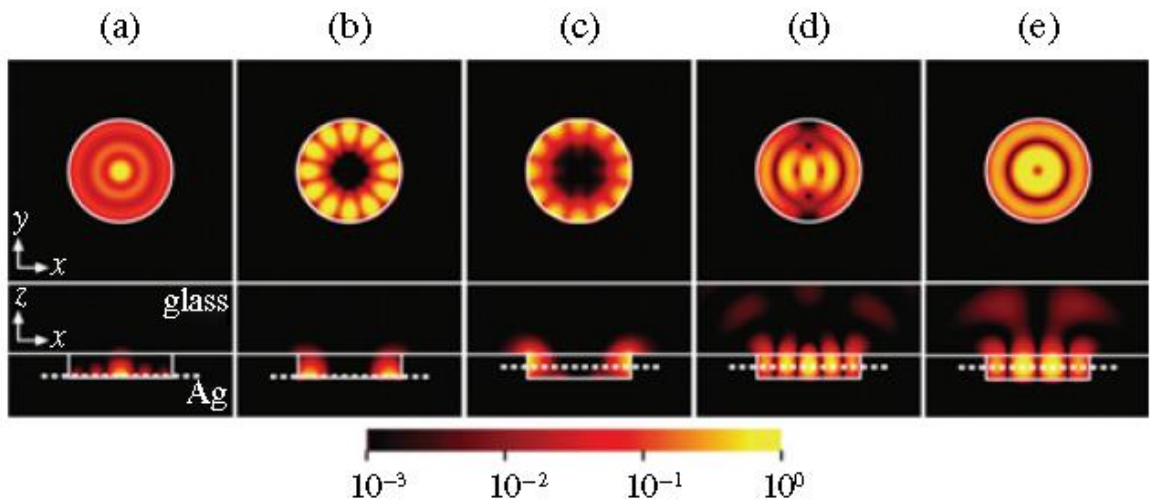

图 1-5 多量子阱圆盘腔半导体-金属纳米激光器的光学模式：(a) 辐射式类 TM 模表面等离极化激元模，(b) 回音壁模类 TM 模表面等离极化激元模，(c) 回音壁模类 TE 模表面等离极化激元模，(d) 偶极式介质模，(e) 单极式介质模 [22,23]。

## 1.3 本文主要研究内容

为进一步实现电注入纳米激光器的微型化，采用基于亚微米厚度薄膜材料的微盘腔器件是可供选择的途径。其中，上述的圆盘腔纳米激光器是较理想的候选者。并且其支持的表面等离极化激元模式具有小的模式体积，是理想的激光模式。但是该模式首先面临着无法室温工作的困难。受到前述纳米线表面等离极化激元激光的启发，我们拟通过在增益介质微盘和金属壳层间引入透明介质层的方法，减小金属损耗，实现表面等离极化激元激光模式的室温激射。

本研究工作致力于开发基于III-V族材料的在室温下工作的可产生表面等离极化激元激光的纳米激光器。我们设计制作了基于 InGaAsP 多量子阱材料和 InGaAs 体材料的半导体-金属复合圆盘腔纳米激光器，对于器件的光学性能进行了数值仿真和实验表征。

本文第二章介绍半导体-金属复合腔纳米激光器的材料选择，结构设计，光学性能的数值仿真及器件制备的工艺流程。第三章介绍对于以 InGaAsP 多量子阱为增益材料的半导体-金属复合圆盘腔纳米激光器的光学性能的实验表征结果。第四章介绍对于以 InGaAs 体材料为增益材料的半导体-金属复合圆盘腔纳米激光器的光学性能的实验表征结果。

# 第二章 半导体-金属复合腔纳米激光器的设计及制备

## 2.1 引言

本研究致力于开发基于 III-V 族半导体增益材料体系，波长在 1550 nm 附近，器件尺寸在三维上均接近或达到亚微米级的半导体-金属复合腔纳米激光器。在前文的叙述中，我们已经了解了在这一领域的相关研究报道。在这一研究方向上，宁存政教授课题组已经实现了微柱腔激光器从低温下脉冲电注入激射，260 K 下连续电注入激射，到室温下连续电注入激射等一系列优秀的研究成果 [17,19,24-28]。本研究需要在此基础上，进一步减小器件的体积。为了实现在三维上达到亚微米级的器件体积，我们选择用微盘腔代替微柱腔，以减小器件高度维度上的尺寸。同时，为了得到更小的模式体积，我们引入金属结构以期望得到表面等离极化激元模式的激射。本章主要介绍半导体-金属微盘腔的结构设计，光学性能仿真和制备方法。

## 2.2 增益材料结构

我们选择在 InP 衬底上外延生长制备的 III-V 族薄膜材料作为激光器的增益材料。我们选择了 InGaAsP 多量子阱材料和 InGaAs 体材料。InGaAsP 多量子阱材料有更高的发光效率，但是由于量子阱区域薄，量子阱数量少，其所能提供的总增益，并不确定会比同样体积的体材料多。InGaAs 体材料虽然发光效率低于多量子阱，但是在同样的总体积下有源区体积更大。

InGaAsP 多量子阱有源区包含 6 层 $In_{0.84}Ga_{0.16}As_{0.66}P_{0.34}$ 势阱，厚 5 nm，增益波长 1550 nm，每两层量子阱之间为 $In_{0.73}Ga_{0.27}As_{0.53}P_{0.47}$ 势垒层，厚 10 nm，增益波长 1200 nm，共 5 层。在多量子阱结构上下各有一层 100 nm 的 $In_{0.78}Ga_{0.22}As_{0.49}P_{0.51}$ 分隔截止层，增益波长 1200 nm。在此外延薄膜最顶层的是 20 nm 的 InP 保护层。InGaAsP 多量子阱薄膜整体结构如表 2.1 所示，光致发光光谱(PL)如图 2-1 所示，可以观察到材料的增益波长在 1550 nm 附近。

InGaAs 体材料的有源区是一层 170 nm 厚的 $In_{0.53}Ga_{0.47}As$ 薄膜，增益波长 1550 nm。在有源区薄膜上下分别有 15 nm 的 P 型和 N 型 InP 轻掺杂层以及 30 nm 的 P 型和 N 型 InP 重掺杂层。加入掺杂层是为了测试这一薄膜结构是否可以支持激射，并以此作为实现电注入激光器的参考依据。在 N 型 InP 重掺杂层与 InP 衬底之间，有一层重掺杂的 InGaAs 薄膜作为微纳加工中的刻蚀介质层。InGaAs 体材料薄膜整体结构如表 2.2 所示，PL 光谱如图 2-2 所示，可以观察到材料的增益波长在 1550 nm 附近，

增益区线宽比 InGaAsP 多量子阱材料宽。

表 2.1 InGaAsP 多量子阱结构

| No. | Material | Thickness (nm) | Purpose |
|---|---|---|---|
| 1 | InP | 20 | cap layer |
| 2 | $In_{0.78}Ga_{0.22}As_{0.49}P_{0.51}$ | 100 | Q 1.2 μm |
| 3 | $In_{0.84}Ga_{0.16}As_{0.66}P_{0.34}$ | 6 | well |
| 4 | $In_{0.73}Ga_{0.27}As_{0.53}P_{0.47}$ | 10 | barrier |
| 5 | $In_{0.84}Ga_{0.16}As_{0.66}P_{0.34}$ | 6 | well |
| 6 | $In_{0.73}Ga_{0.27}As_{0.53}P_{0.47}$ | 10 | barrier |
| 7 | $In_{0.84}Ga_{0.16}As_{0.66}P_{0.34}$ | 6 | well |
| 8 | $In_{0.73}Ga_{0.27}As_{0.53}P_{0.47}$ | 10 | barrier |
| 9 | $In_{0.84}Ga_{0.16}As_{0.66}P_{0.34}$ | 6 | well |
| 10 | $In_{0.73}Ga_{0.27}As_{0.53}P_{0.47}$ | 10 | barrier |
| 11 | $In_{0.84}Ga_{0.16}As_{0.66}P_{0.34}$ | 6 | well |
| 12 | $In_{0.73}Ga_{0.27}As_{0.53}P_{0.47}$ | 10 | barrier |
| 13 | $In_{0.84}Ga_{0.16}As_{0.66}P_{0.34}$ | 6 | well |
| 14 | $In_{0.78}Ga_{0.22}As_{0.49}P_{0.51}$ | 100 | Q 1.2 μm |
| 15 | InP | … | substrate |

表 2.2 InGaAs 体材料结构

| No. | Material | Thickness (nm) | Doping ($cm^{-3}$) | Purpose |
|---|---|---|---|---|
| 0 | p-InP | 30 | $>10^{18}$ | p-contact |
| 1 | p-InP | 15 | $10^{18}$ | Light doped |
| 2 | $In_{0.53}Ga_{0.47}As$ | 170 | no doping | Gain layer |
| 3 | n-InP | 15 | $10^{18}$ | Light doped |
| 4 | n-InP | 30 | $>10^{18}$ | n-contact |
| 5 | $In_{0.53}Ga_{0.47}As$ | 50 | $>10^{18}$ | Etch stop |
| 6 | n-InP | … | … | Substrate |

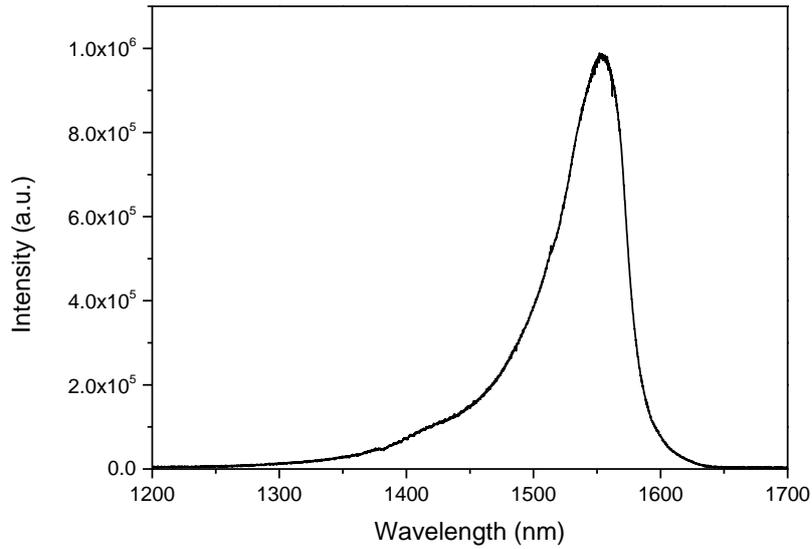

图 2-1 InGaAsP 多量子阱 PL 光谱

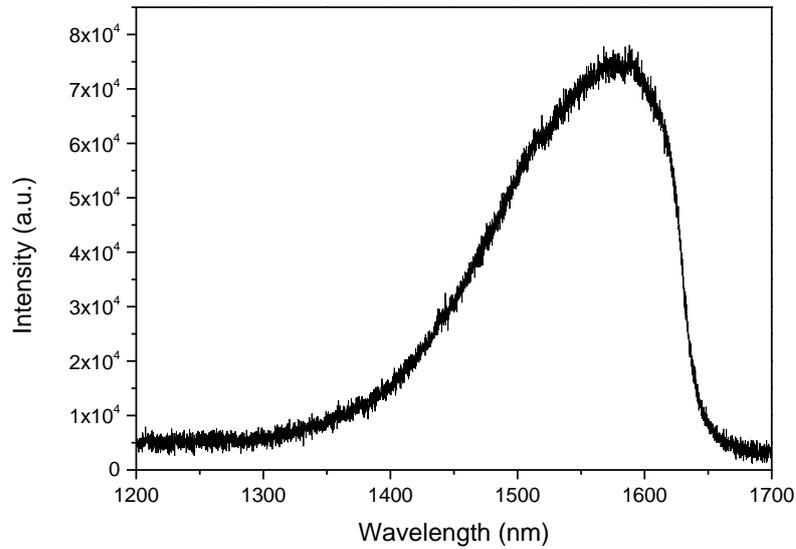

图 2-2 InGaAs 体材料 PL 光谱

## 2.3 半导体-金属复合圆盘腔纳米激光器结构设计与性能仿真

本研究中的激光器是半导体-金属复合圆盘微腔结构。有关这一类器件已有一些研究报道。这一结构的基本组合方式是增益介质圆盘和包覆圆盘底面和侧壁的金属薄膜。该类激光器一般有三种模式：类 TM 模表面等离极化激元模 (TM-like SPP)、类 TE 模表面等离极化激元模 (TE-like SPP) 和 TE 介质模 [22,23]。对于增益介质和金属膜直接接触的器件，相

关研究结果指出，在室温下由于金属带来的光学的损耗，两种 SPP 模均无法激射。为了减小金属引入的损耗，提高器件的工作温度，我们在介质圆盘和金属包覆层之间加入一层低折射率的透明介质层。这样可以将 SPP 模式导入透明介质层，有效的减小金属损耗，提高器件的工作温度。图 2-3 为 InGaAsP 多量子阱圆盘腔半导体-金属复合腔纳米激光器的结构示意图。该器件包含一个多量子阱圆盘，一层 $Al_2O_3$ 透明介质层和一层银薄膜。

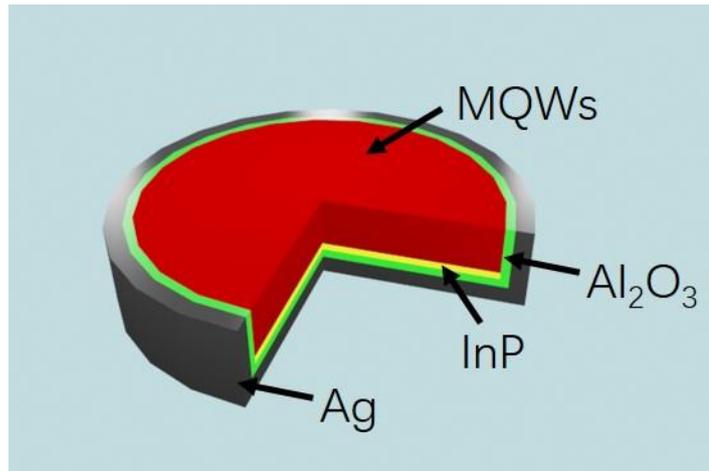

图 2-3 InGaAsP 多量子阱圆盘腔半导体-金属复合腔纳米激光器

图 2-4 是一个介质微盘直径 600 nm 的 InGaAsP 多量子阱材料复合圆盘腔纳米激光器的 TM-like SPP 模、TE-like SPP 模和 TE 介质模中最具代表性的模式电场分布的数值仿真 (COMSOL) 结果。图 2-4-a 是 $TM_{11}$ SPP 模式，我们看到电场强度最强的区域分布在介质圆盘底面与金属包覆层之间的透明介质层中，而电位移矢量方向垂直于金属与介质的交界面，表明该模式为 SPP 模式，形成于介质圆盘底面与金属的交界面。图 2-4-b 是 $TE_{31}$ SPP 模式，我们看到电场强度最强的区域分布在介质圆盘侧壁与金属包覆层之间的透明介质层中，而电位移矢量方向垂直于金属与介质的交界面，表明该模式为 SPP 模式，形成于介质圆盘侧壁与金属的交界面。图 2-4-c 是 $TE_{01}$ 介质模式，我们看到电场强度最强的区域分布在介质圆盘内，而电位移矢量方向在介质圆盘内形成闭环，表明该模式为介质模式。

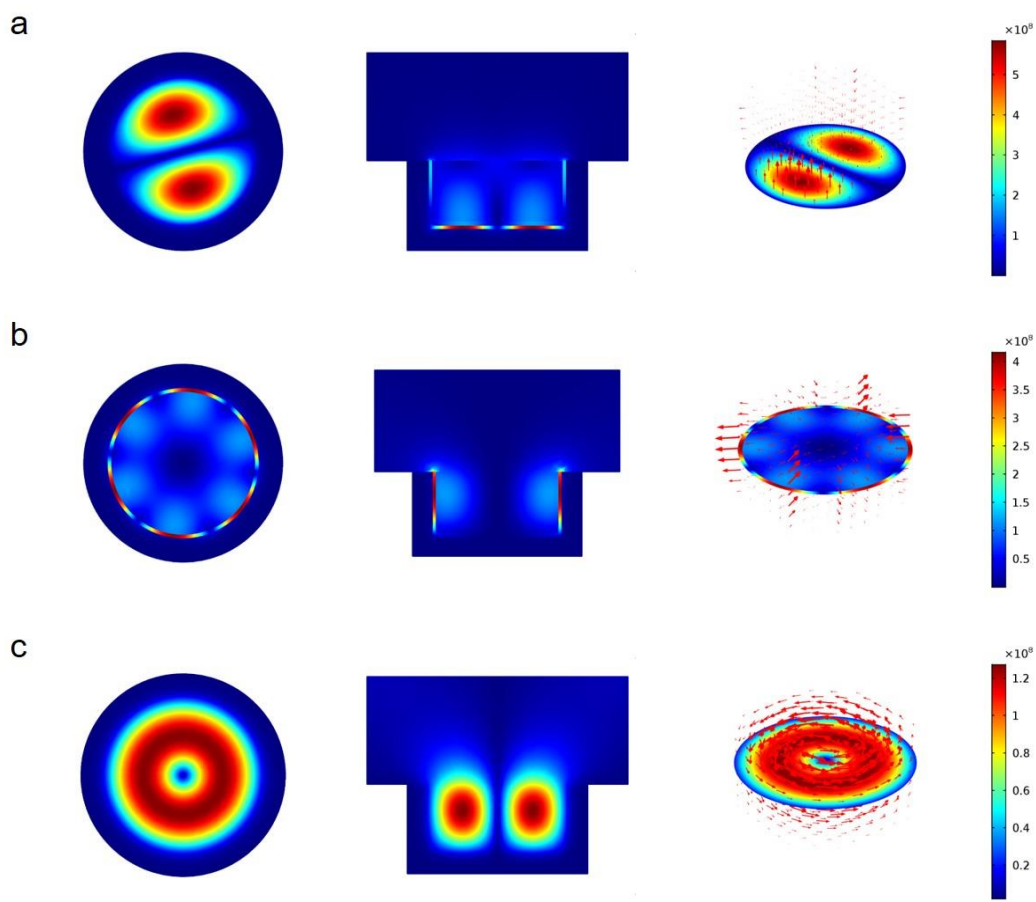

图 2-4 介质微盘直径 600 nm 的 InGaAsP 多量子阱半导体-金属复合圆盘腔纳米激光器典型光学模式：(a) 类 TM 表面等离极化激元模, (b) 类 TE 表面等离极化激元模, (c) TE 介质模。

在半导体-金属复合圆盘腔中，对于器件光学性能影响最主要的两个影响因素是半导体圆盘直径和透明介质层的厚度。图 2-5 展示了透明介质层厚度为 10 nm 的圆盘腔激光器各主要模式的波长与圆盘直径的相互关系。可以发现，随圆盘直径增加，同一模式的波长明显红移。图 2-6 展示了同样的透明介质层厚度的器件各主要模式的模式体积与半波长极限 $(\lambda/2n)^3$ 的比值和圆盘直径的相互关系。可以发现，随着圆盘直径的增加，同一模式的模式体积相对于所涉及波长的半波长极限的比值逐渐减小。值得注意的是，所有的 SPP 模式都具有明显小于介质模式的比值，尤其是低阶的 SPP 模式，其比值甚至小于 0.5。这充分体现了 SPP 模式相对于介质模式更加优秀的场限制能力。

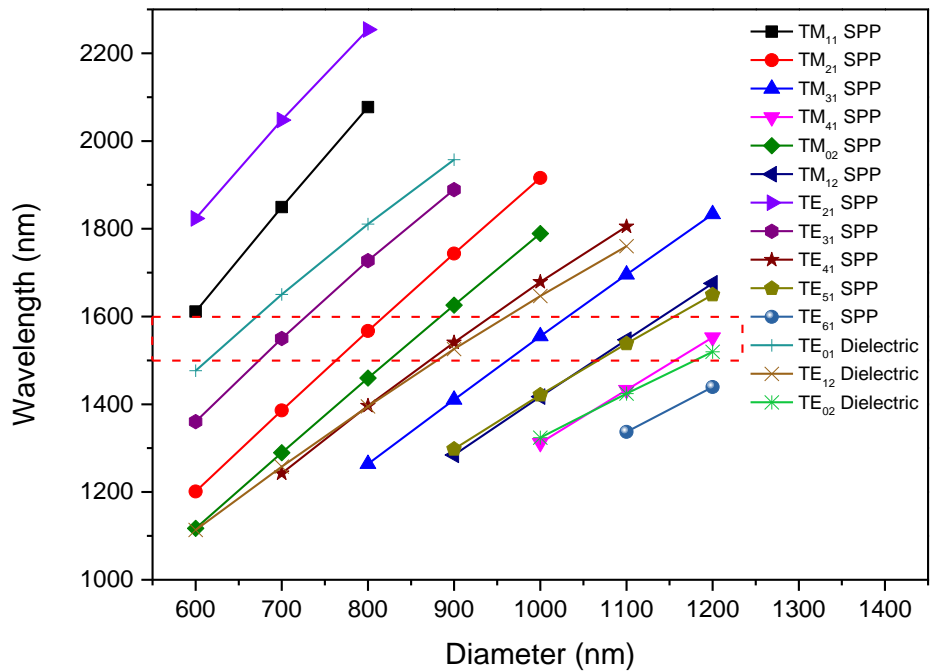

图 2-5 透明介质层厚度为 10 nm 的圆盘腔激光器各主要模式的波长与圆盘直径的相互关系。红色虚线方框为增益材料增益波长范围。

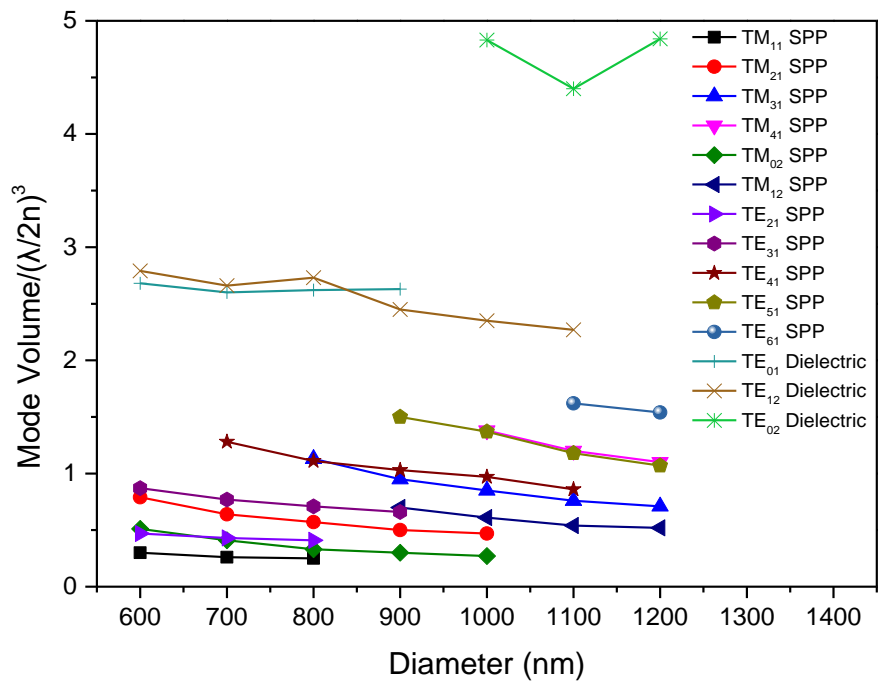

图 2-6 透明介质层厚度 10 nm 的器件各主要模式的模式体积与半波长极限$(\lambda/2n)^3$的比值和圆盘直径的相互关系。

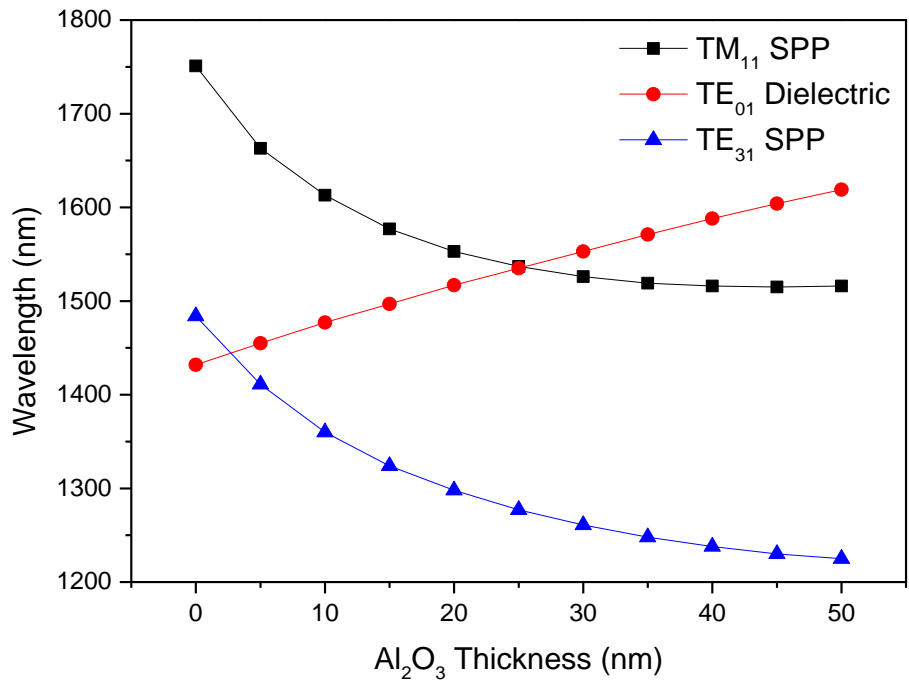

图 2-7 圆盘直径 600 nm 的器件 $TM_{11}$ SPP 模式、$TE_{01}$ 介质模式和 $TE_{31}$ SPP 模式的波长与透明介质层厚度之间的关系。

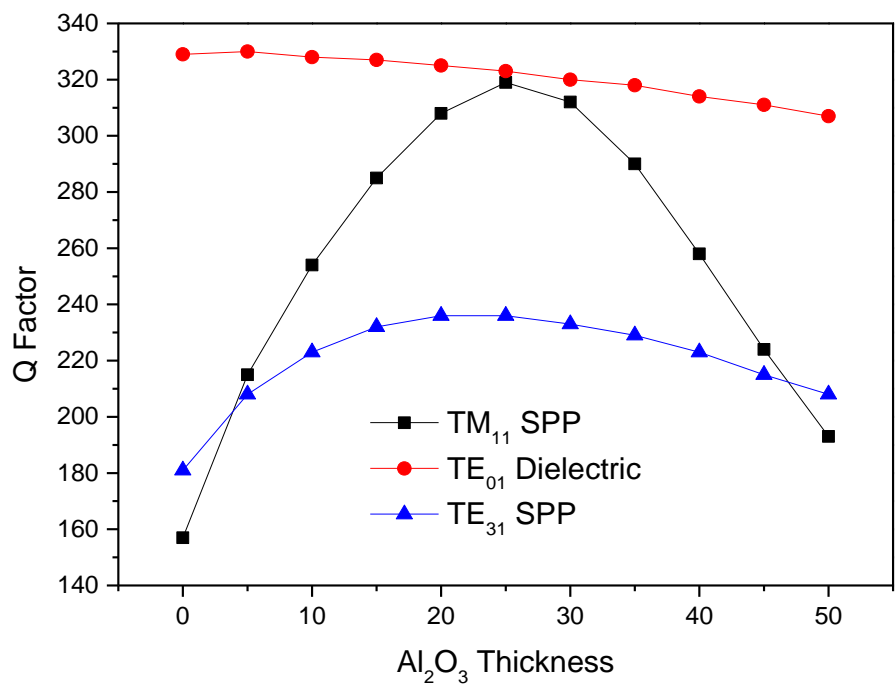

图 2-8 圆盘直径 600 nm 的器件 $TM_{11}$ SPP 模式、$TE_{01}$ 介质模式和 $TE_{31}$ SPP 模式品质因子与透明介质层厚度之间的关系。

图 2-7 和图 2-8 为一个圆盘直径 600 nm 的器件三个主要模式 ($TM_{11}$ SPP 模式、$TE_{01}$ 介质模式和 $TE_{31}$ SPP 模式) 的波长和品质因子与透明介质层厚度之间的关系。从图中我们可以看出，随着介质层厚度增加，同一介质模式的波长发生了一个近似匀速的红移，这可以解释为因为透明介质层材料折射率小于增益介质，微腔中模式所处区域的等效折射率随介质层厚度增加而减小。而两种 SPP 模式的波长随着透明介质层厚度的增加表现出了一个先快速蓝移，进而转变为缓慢蓝移的行为，这同样与等效折射率有关。由于透明介质层的折射率高于金属膜，对于 SPP 模式来说，引入透明介质层，将 SPP 模式导入介质层，等效折射率明显提高导致波长蓝移。而就品质因子的变化而言，介质模式的品质因子随透明介质层厚度的增加表现出缓慢减小的趋势，但是在 0~50 nm 范围内，品质因子始终保持在 300 以上。而两种 SPP 模式的品质因子则随着透明介质层的厚度增加展现出了先快速增加，之后快速下降的变化。尤其是 $TM_{11}$ SPP 模在介质层厚度由 0 增加到 25 nm 的过程中，品质因子迅速由 150 左右增加至接近 320，之后又快速下降，在介质层厚度达到 50 nm 时品质因子降到不足 200。这是因为在透明介质层很薄时，SPP 模式被导入介质层，减小了金属造成的损耗，模式的品质因子得到了有效的提高。而当介质层厚度太厚时，由于金属与增益介质距离太大，SPP 模式不易形成，因此模式的品质因子转而下降。参考以上影响因素的仿真结果，我们选择透明介质层的厚度范围在 10~15 nm。

InGaAs 体材料有源区的半导体-金属微盘腔纳米激光器的器件设计基于和 InGaAsP 多量子阱的器件同样的设计原理，并有同样种类的模式和同样与之相关的影响因素。由于体材料的效率相比多量子阱材料低，我们希望在尽量减小金属对器件性能的损耗的同时依然可以得到 SPP 模式的激射，因此我们选择保留 TM-like 或 TE-like 两种 SPP 模式中的一种。考虑到微纳加工的可操作性和难易度，以及模式分布与有源区重叠的区域面积，我们选择保留分布在侧壁的 TE-like SPP 模式。在器件设计上，我们在 InGaAs 体材料薄膜圆盘的底面与 $Al_2O_3$ 透明介质层和 Ag 薄膜间添加一层 100 nm 左右的 $SiO_2$ 隔离层。在文献报道中，这一厚度已足够阻止 SPP 模式的产生和减小金属损耗 [20]。图 2-9 为 InGaAs 体材料圆盘腔半导体-金属复合腔纳米激光器的结构示意图。该器件包含一个体材料圆盘，一层 $SiO_2$ 隔离层，一层 $Al_2O_3$ 透明介质层和一层银薄膜。

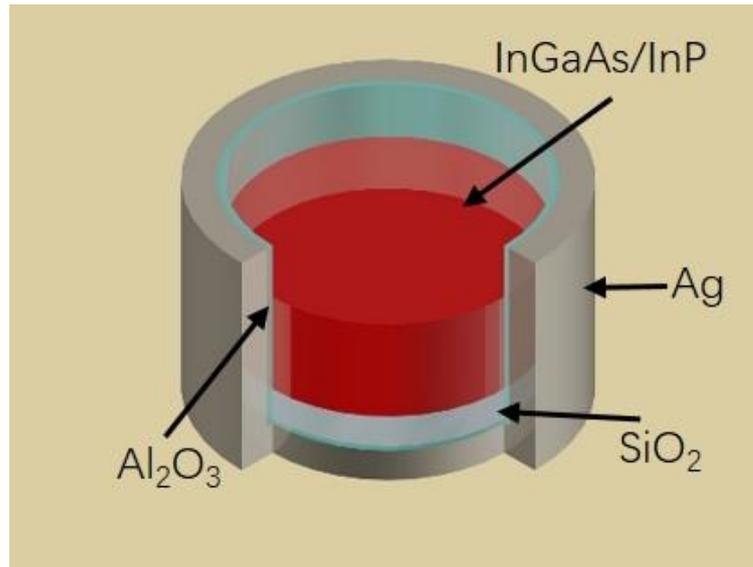

图 2-9 InGaAs 体材料圆盘腔半导体-金属复合腔纳米激光器

**2.4 半导体-金属复合圆盘腔纳米激光器制备工艺**

  基于多量子阱材料和体材料的半导体-金属复合腔纳米激光器均通过标准化的微纳加工工艺批量制备。第一步，200 nm 的 $SiO_2$ 薄膜被沉积在外延生长增益材料薄膜表面，作为干法刻蚀的硬掩膜层。第二步，依照设计尺寸的纳米圆盘阵列通过电子束曝光负性电子束抗蚀剂在 $SiO_2$ 薄膜上画出并显影。第三步，采用干法刻蚀工艺先后将圆盘图形转移到 $SiO_2$ 薄膜层和外延薄膜层。对于多量子阱薄膜材料的器件剩下的 $SiO_2$ 掩模会被氢氟酸洗去，而对于体材料薄膜的器件，$SiO_2$ 掩模将会被保留作为隔离层。第四步，15 nm 的 $Al_2O_3$ 将会采用原子层沉积系统沉积于干法刻蚀而成的结构表面。第五步，100 nm 的 Ag 薄膜通过磁控溅射沉积于 $Al_2O_3$ 薄膜外。第六步，全部器件连同 InP 衬底翻转后键合在 Si 衬底上，Ag 薄膜与 Si 衬底间以键合胶链接。第七步，用浓盐酸对 InP 进行湿法刻蚀，待 InP 衬底全部被刻蚀掉后，器件制备完成。图 2-10 为器件制备流程示意图，以基于 InGaAsP 多量子阱材料的纳米激光器为例。

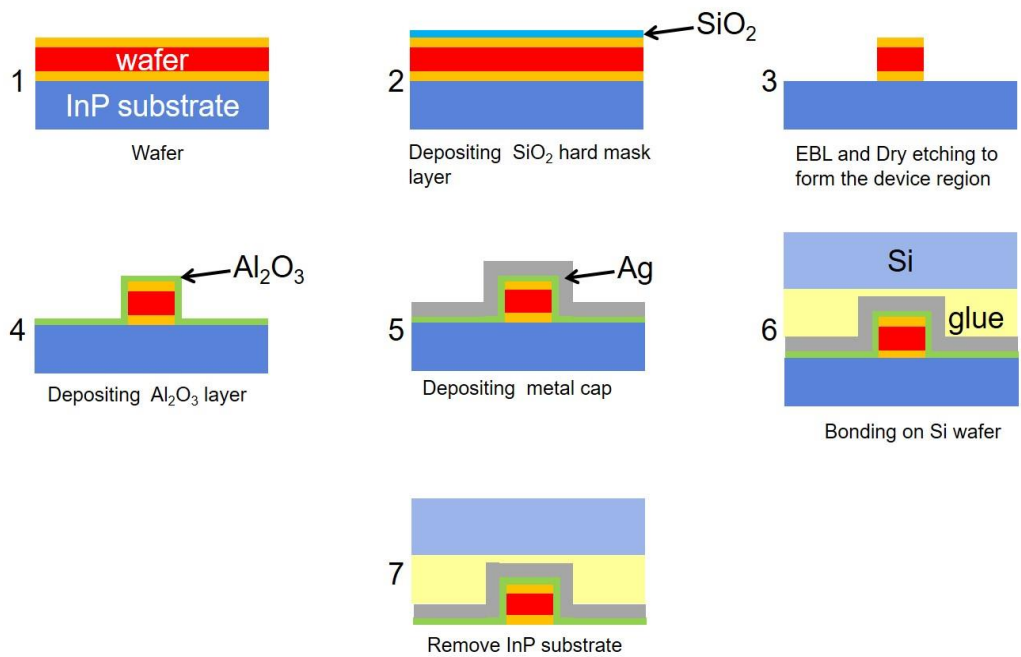

图 2-10 InGaAsP 多量子阱半导体-金属复合圆盘腔纳米激光器制备流程

## 2.5 本章小结

  本章中，我们介绍了基于 III-V 族材增益料体系的半导体-金属复合腔纳米激光器的材料结构，器件设计，光学性能仿真和器件制备工艺。我们选择了 InGaAsP 多量子阱薄膜和 InGaAs 体材料薄膜作为激光器的增益材料并设计了改进的半导体-金属复合圆盘腔纳米激光器，通过在有源区和金属壳层间加入一层薄层透明介质层将 SPP 模式耦合进介质层，有效减小 SPP 带来的损耗，提高 SPP 模式的品质因子。通过有限元数值仿真，我们发现随半导体圆盘的直径增加，该类器件各个光学模式的波长而红移，同时其模式体积与 $(\lambda/2n)^3$ 的比值则逐渐减小，并且表面等离极化激元模式具有明显小于介质模式的模式体积。另一方面，随着透明介质层厚度增加，同一直径的器件介质模式的波长会发生红移而 SPP 模式的波长会发生蓝移，介质模式的品质因子变化平稳而 SPP 模式的品质因子会先急速提高后急速下降。对于 600 nm 直径的器件，10 至 15 nm 的介质层厚度适宜 SPP 模式激射。本研究中的半导体-金属复合腔纳米激光器由标准化的半导体微纳加工工艺批量制备。

# 第三章 InGaAsP 多量子阱半导体-金属腔纳米激光器性能表征

## 3.1 引言

在完成器件的设计和制备后,对于器件性能的测试表征随之展开。本章对基于 InGaAsP 多量子阱增益材料体系的半导体-金属复合圆盘腔纳米激光器进行了器件形貌和光学性能的表征。通过扫描电子显微镜 (SEM) 对器件的形貌进行了观察。通过光致发光 (PL) 测试系统对多个规格的器件在室温下光泵浦激射的性能进行了表征。表征结果证实我们得到了该类器件室温下 SPP 模式的激光发射。

## 3.2 InGaAsP 多量子阱纳米激光器形貌表征

我们在器件制备过程中对样品进行扫描电子显微镜观察,以确定器件制备的效果。图 3-1 为直径 610 nm 的器件在干法刻蚀并去除残留掩模后,磁控溅射沉积 Ag 薄膜后和器件制备完成后的扫描电子显微镜照片。从干法刻蚀后的圆盘形貌照片可以观察到多量子阱结构以被刻蚀成圆盘形状,圆盘侧壁较平滑,但是圆形并不完美。这是由电子束曝光的仪器误差造成的。从完成 Ag 薄膜沉积的结构的形貌照片可以观察到银薄膜已经完全覆盖了多量子阱圆盘的表面。Ag 薄膜是颗粒沉积而成,颗粒粒径较小,堆积紧密。从制备完成的器件的形貌照片可以观察到 InP 衬底已经被完全刻蚀掉,露出多量子阱圆盘表面。器件的形状从照片给出的俯视角度观察并不是完美的圆形。

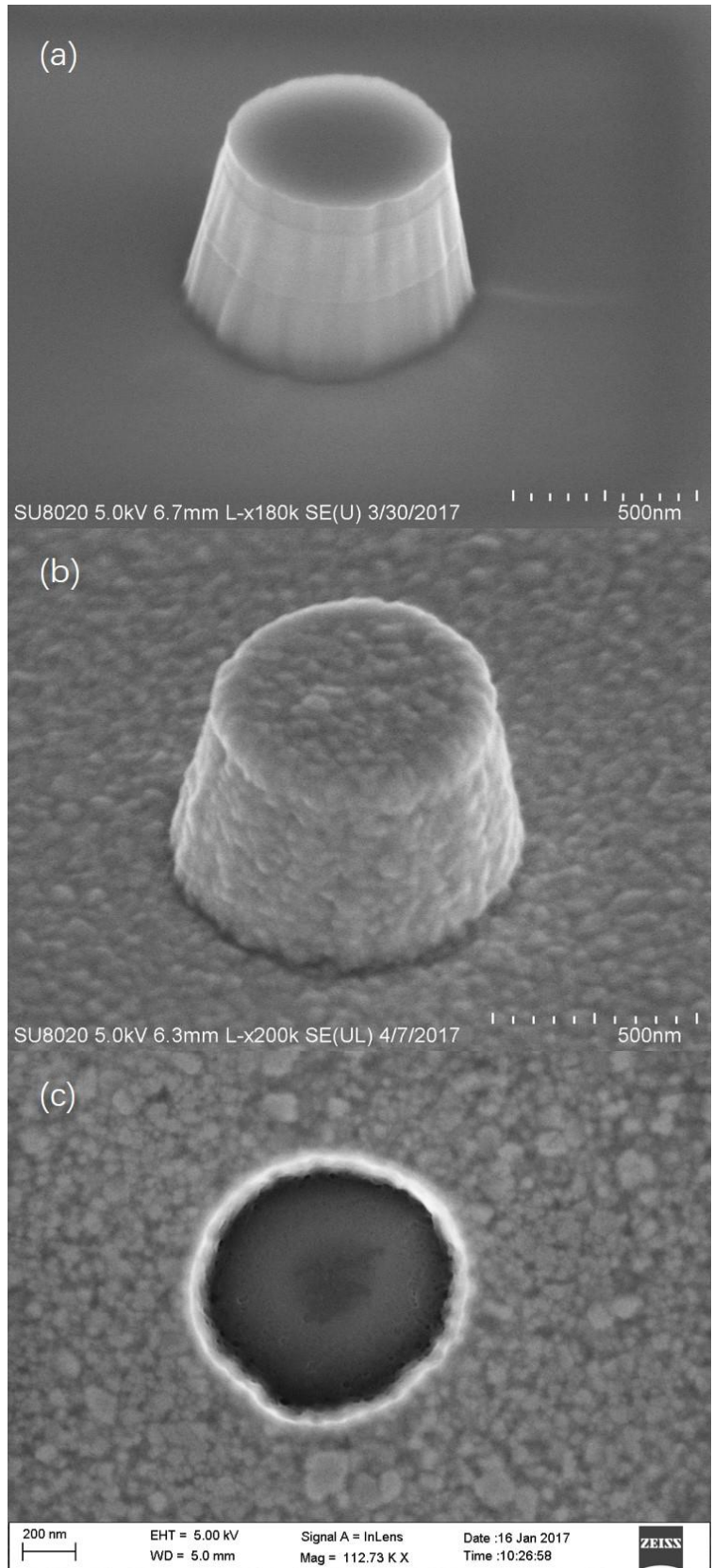

图 3-1 直径 610 nm 器件在 SEM 照片：(a) 干法刻蚀并去除残留掩模后，(b) 磁控溅射沉积 Ag 薄膜后，(c)器件制备完成后。

## 3.3 InGaAsP 多量子阱纳米激光器光学性能表征

基于 InGaAsP 多量子阱增益材料体系的纳米激光器的光学性能测试均在室温下测试。测试系统是实验室自行搭建的 PL 测试系统，器件通过光泵浦激发。对于我们的器件，泵浦光源是波长 970 nm，频率 80 MHz 的皮秒脉冲激光。测试中，泵浦光由一个 100 倍物镜聚焦在被测试器件上。器件被激发后发出的光由同一个物镜收集，经过滤光片过滤掉激发光源发光后导入光谱仪，得到器件的发射光谱。在系统中加入偏振片，还可以对光学模式的偏振进行测试。

图 3-2 展示了一个直径 610 nm 器件的 PL 光谱随泵浦光功率提高的演化过程。可以看到，随着泵浦功率逐渐提高，在 1590 nm 附近出现了一个尖锐的发光峰，并且该发光峰的强度表现出比背景光更快速地增长。在泵浦功率为 0.5 µW 时，此发光峰强度仅略高于背景，而泵浦功率为 22.5 µW 时，此发光峰已明显高于背景，表明在此波长有一个光学模式存在且该模式可能为激光模式。

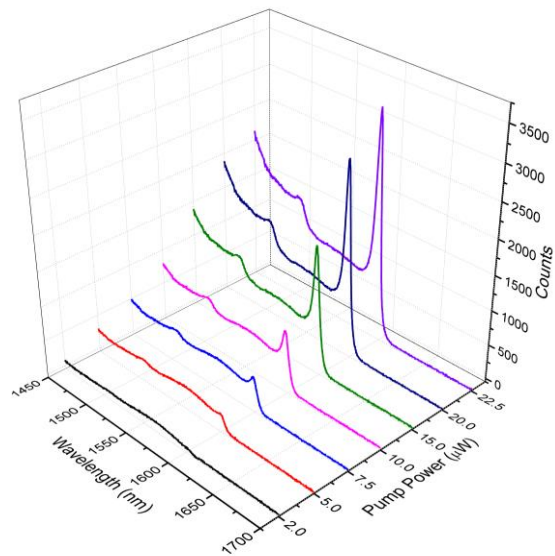

图 3-2 直径 610 nm 器件的 PL 光谱演化

图 3-3 为图 3-2 中光学模式的输入输出曲线。可以看出，双对数坐标系中由数据点组成的曲线程 S 形状，表现出典型的激射特征。在低泵浦和高泵浦下，曲线斜率约等于 1，呈线性，对应于自发辐射阶段和受激辐射阶段，连接这两个阶段的曲线斜率大于 1，程非线性，对应于自发辐射放大阶段，激光阈值约为 10 µW。彩色曲线是对应于实验结果计算的激光速率方程曲线。实验数据点曲线对应于自发辐射耦合因子为 0.15 的曲线。

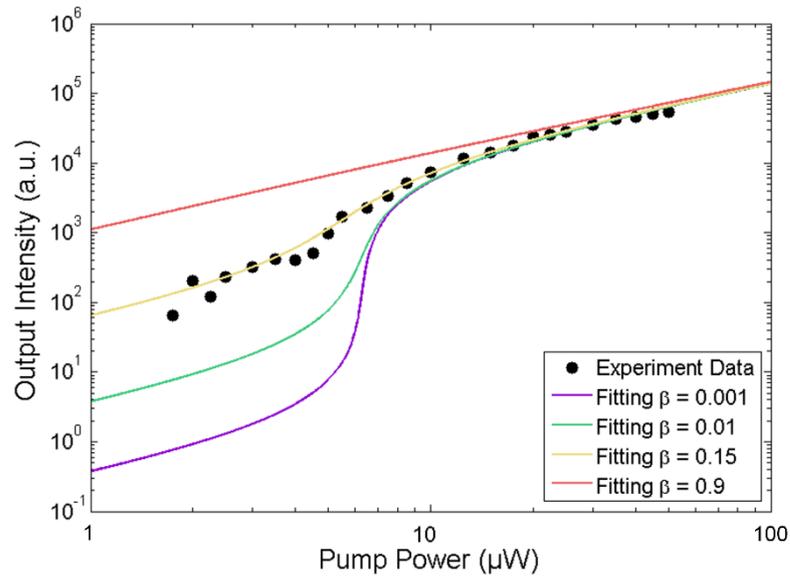

图 3-3 直径 610 nm 器件 1590 nm 光学模式的输入输出曲线

图 3-4 为图 3-2 中光学模式发光峰的线宽与泵浦功率的关系曲线。在泵浦功率大于 5 μW 后,发光峰的线宽开始随着泵浦功率提高减小,符合激射的特点。在泵浦功率大于 25μW 后,线宽随着泵浦功率提高缓慢增加,这可以解释为激光的热效应导致线宽增加。对应于 25μW 泵浦功率的线宽为 7.4 nm,此时光学模式的波长为 1590 nm,计算可得到其品质因子为 215。

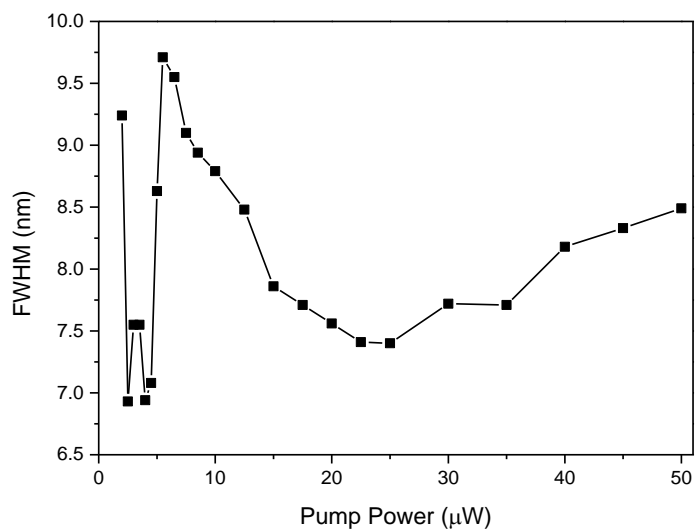

图 3-4 直径 610 nm 器件 1590 nm 模式的线宽与泵浦功率的关系曲线

图 3-5 为图 3-2 中光学模式发光峰的发光强度的倒数与线宽的关系曲线。可以观察到在泵浦功率超过激光阈值至发光峰线宽因热效应展宽前，线宽和发光强度的倒数呈线性关系，符合激光的 Schawlow-Townes 方程。结合图 3-2 至图 3-5 所呈现的结果，我们认为该光学模式是一个激光模式。

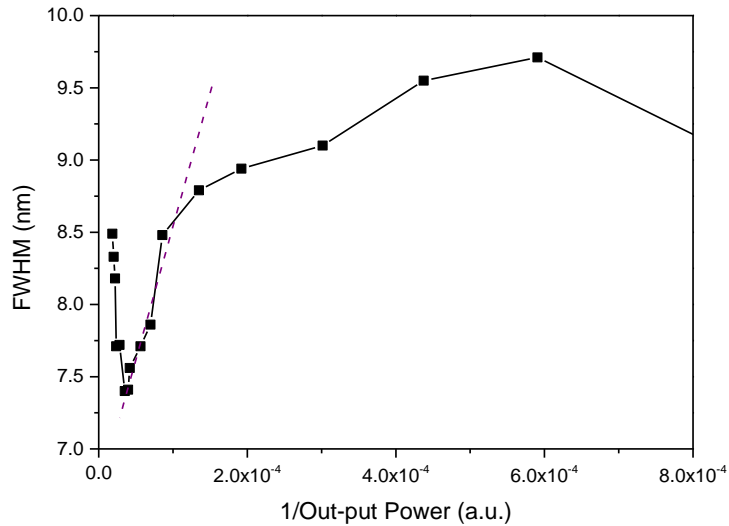

图 3-5 直径 610 nm 器件 1590 nm 模式发光强度倒数与线宽的关系曲线

在确定这个 610 nm 器件在 1590 nm 的发光模式为激光后，我们要确定这一模式的种类。对照前述器件数值仿真的结果，这一尺寸的器件在此波长的模式应该为 $TM_{11}$ SPP 模式。为进一步验证这一激光模式的性质，我们表征了这一模式的偏振特性。图 3-6 为该光学模式的偏振测试结果，可以确定这一模式具有典型的线偏振。图 3-7 为 $TM_{11}$ SPP 模式，$TE_{31}$ SPP 模式和 $TE_{01}$ 介质模式的远场的电场强度和电位移矢量分布的数值仿真结果。可以发现，只有 $TM_{11}$ SPP 模式为明显的线偏振，其他两个模式均不能得到线偏振的测试结果。因此我们可以确定我们在实验中测得的激光模式为 TM-like SPP 模式。我们得到了器件尺寸为亚微米级的半导体-金属圆盘腔激光器室温下的 SPP 模式的激光发射。

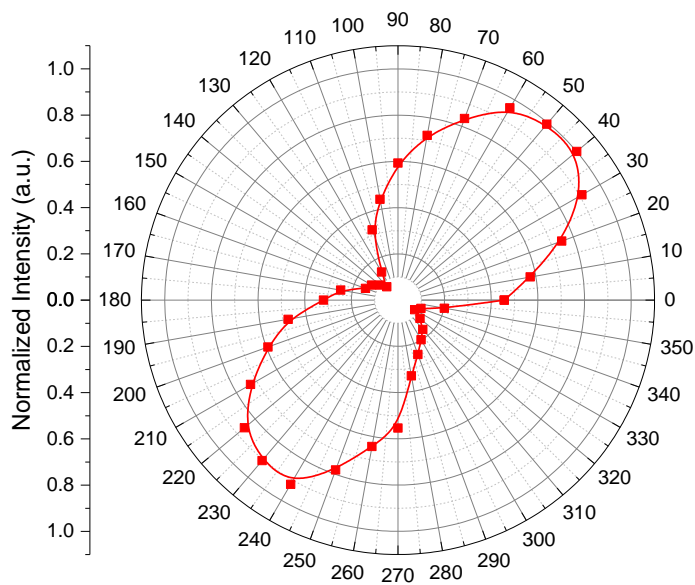

图 3-6 直径 610 nm 器件 1590 nm 模式的偏振

对于该类器件的进一步研究发现该类器件可以同时支持多个激光模式。图 3-8 为一个直径为 1 μm 的激光器在泵浦光功率为 100 μW、350 μW 和 430 μW 时的 PL 光谱。可以观察到，在泵浦功率为 100 μW 时，只在 1550 nm 波长附近有一个激光模式，而在 350 μW 的泵浦功率下，在 1310 nm 波长附近出现了另一个激光模式，当泵浦光功率为 430 μW 时，在 1180 nm 波长附近，出现了第三个激光模式。图 3-9 为另一个具有同样三个激光模式的 1 μm 激光器的三个激光模式的输入输出曲线。可以发现三个模式是随着泵浦功率的提高先后出现的，且三个模式的波长原来越短，表明模式对应的光子能量逐渐增高。我们分析认为这是因为小尺寸器件量子阱的体积有限，导致量子阱的势阱能带随泵浦功率提高被载流子填充饱和，所以量子阱子带和势垒被载流子填充，产生了对应于这两个能级的新的增益。因此，出现在 1310 nm 和 1180 nm 的两个激光模式和增益很可能来自于量子阱子带和势垒。对于不同体积的多量子阱材料 PL 测试证实了这一猜想。图 3-10 为经干法刻蚀后的直径 7 μm 和直径 2 μm 的圆盘结构的归一化 PL 光谱随泵浦功率提高的演化过程。从图中可以观察到直径 2 μm 圆盘结构的 PL 光谱形状随泵浦功率的提高发生了很大的变化。在泵浦功率由 0 提高到 1.65 mW 的过程中先后在 1450 nm、1300 nm 和 1200 nm 附近出现了新的发光峰，对应于量子阱子带和势垒增益的能级。直径 7 μm 圆盘结构的 PL 光谱形状在泵浦功率达到 1.65 mW 时仍未发生明显的变化，接近未经加工的多量子阱薄膜。这一测试表明对于体积极小的 InGaAsP 多量子阱材料，在高泵浦下势阱能级将会出现饱和，而势阱子带

和势垒能级将会被激发产生增益。因此对于接近或亚微米级的 InGaAsP 多量子阱激光器，在高泵浦下会观察到对应于势阱子带和势垒增益的激光。

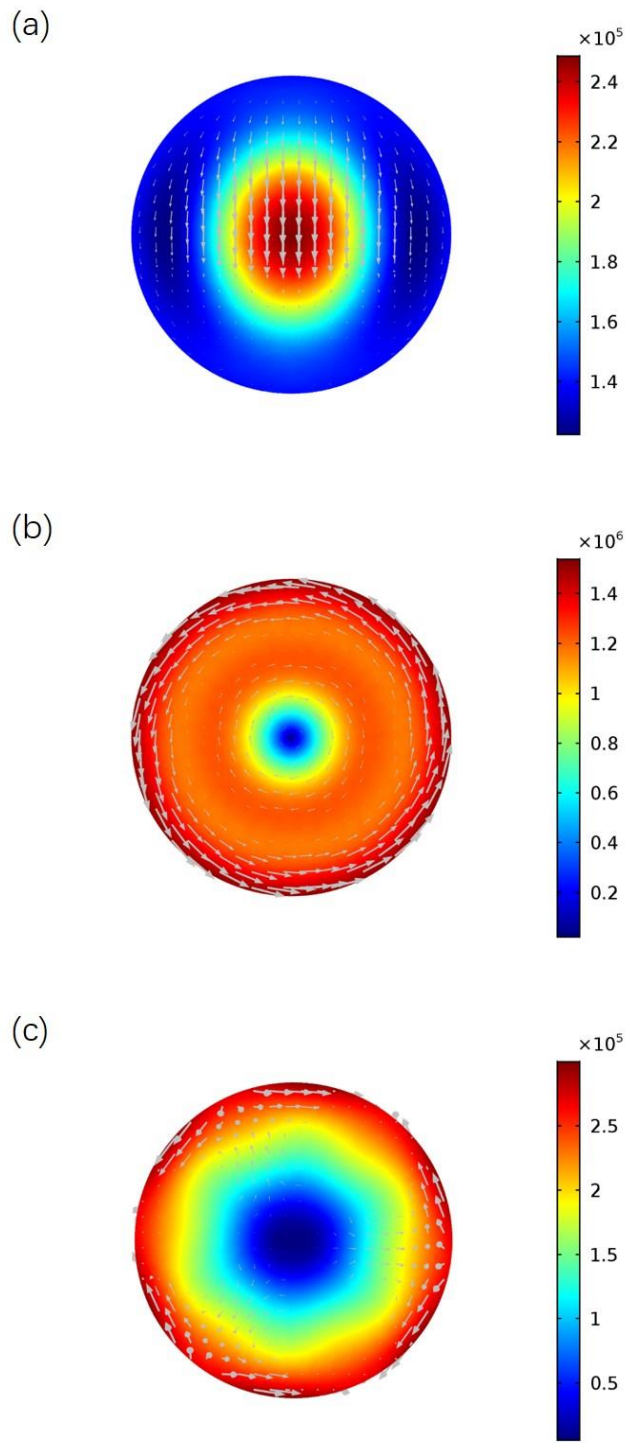

图 3-6 直径 610 nm 器件 (a) $TM_{11}$ SPP 模式，(b) $TE_{31}$ SPP 模式，(c) $TE_{01}$ 介质模式的远场的电场强度和电位移矢量分布的数值仿真

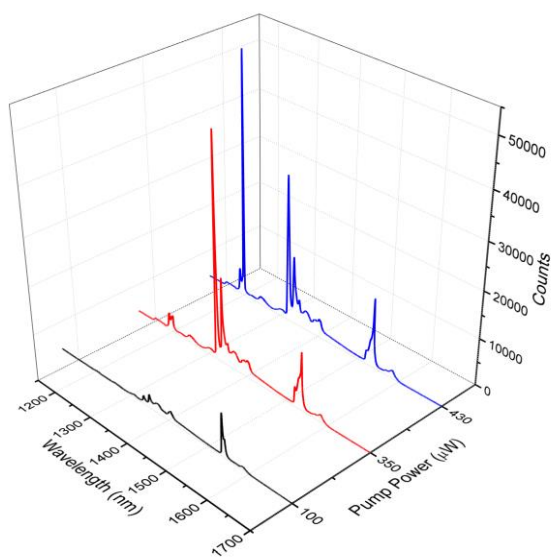

图 3-8 直径 1 μm 的激光器在泵浦功率 100 μW、350 μW 和 430 μW 时的 PL 光谱

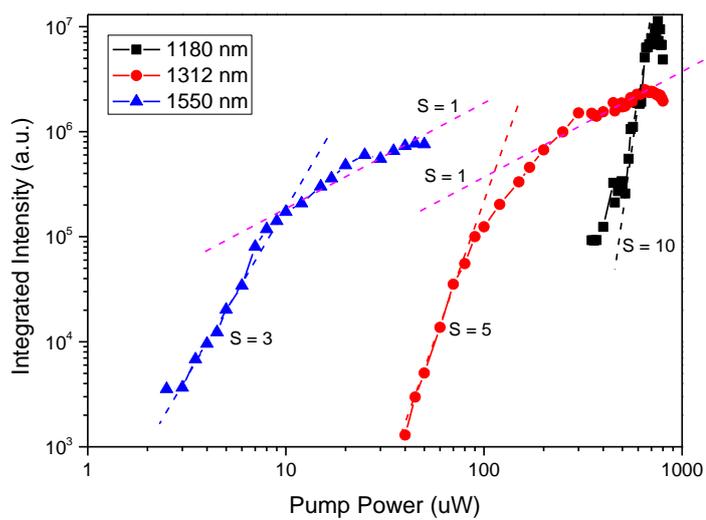

图 3-9 直径 1 μm 激光器 1180 nm、1312 nm 和 1550 nm 激光模式的输入输出曲线

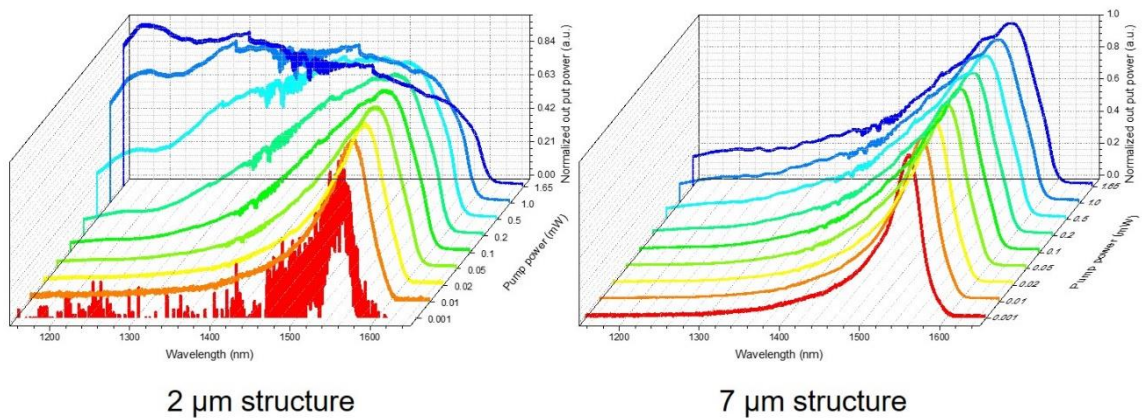

图 3-10 干法刻蚀后的直径 2 μm (左)和直径 7 μm (右)的圆盘结构的归一化 PL 光谱演化过程

### 3.4 本章小结

本章介绍了以 InGaAsP 多量子阱为增益材料的半导体-金属复合圆盘腔纳米激光器的实验表征。在形貌表征中我们发现器件形貌较好，但是电子束曝光带来的系统误差无法避免。而在室温下的光学表征中，我们实验证实了半导体圆盘腔直径为 600 nm 的器件的室温下 $TM_{11}$ SPP 模式的激射，为表面等离极化激元模式激光。此外我们发现，由于极小的器件体积，在高泵浦功率下，对应于 1550 nm 波长的多量子阱势阱能级被填充饱和，载流子会填充势阱的子带能级和势垒能级，产生与之对应的增益，支持对应波长的激射。对于圆盘直径 1 μm 的器件，可以依次产生波长 1550 nm、1310 nm 和 1180 nm 的激光。

# 第四章 InGaAs 体材料半导体-金属腔纳米激光器性能表征

## 4.1 引言

相比于 InGaAsP 多量子阱材料，InGaAs 体材料虽然效率低，但是同样体积的器件中有源区体积大于 InGaAsP 多量子阱，因此也能提供较好的增益。而且 InGaAs 体材料不同于 InGaAsP 多量子阱材料存在势阱，势阱子带和势垒，只有体材料的能带结构。因此，基于体材料的半导体-金属纳米激光器应该会具有不同于基于多量子阱材料的纳米激光器的器件性能。本章对基于 InGaAs 体材料的半导体-金属复合圆盘腔纳米激光器进行了器件形貌和光学性能的表征。表征方法同样是扫描电子显微镜观察和 PL 测试。测试发现对于 InGaAs 体材料作为有源区的半导体-金属纳米激光器，可通过光泵浦得到室温下的激光发射。在 77 K 的低温条件下，直径 400 nm 的纳米激光器可以在光泵浦下产生激射。

## 4.2 InGaAs 体材料纳米激光器形貌表征

图 4-1 为直径 900 nm 左右的器件在干法刻蚀并去除残留掩模后，磁控溅射沉积 Ag 薄膜后和器件制备完成后的扫描电子显微镜照片。从干法刻蚀后的圆盘形貌可以发现，InGaAs 体材料薄膜已被刻蚀成圆盘状，圆盘侧壁较平滑，但是圆形并不完美。同时，圆盘的侧壁不够陡直，这是由于电子束曝光艺中，电子束在电子束抗蚀剂中发生了散射，导致被曝光的抗蚀剂形成的圆盘具有了不陡直的侧壁，因此这一形貌在干法刻蚀工艺后也被转移到了 InGaAs 圆盘。完成 Ag 薄膜沉积的结构的形貌照片显示 Ag 薄膜已经完全覆盖了 InGaAs 体材料圆盘的表面。组成 Ag 薄膜的颗粒粒径较小，堆积紧密。从制备完成的器件的形貌照片可以观察到 InP 衬底已经被完全刻蚀掉，露出体材料圆盘表面。

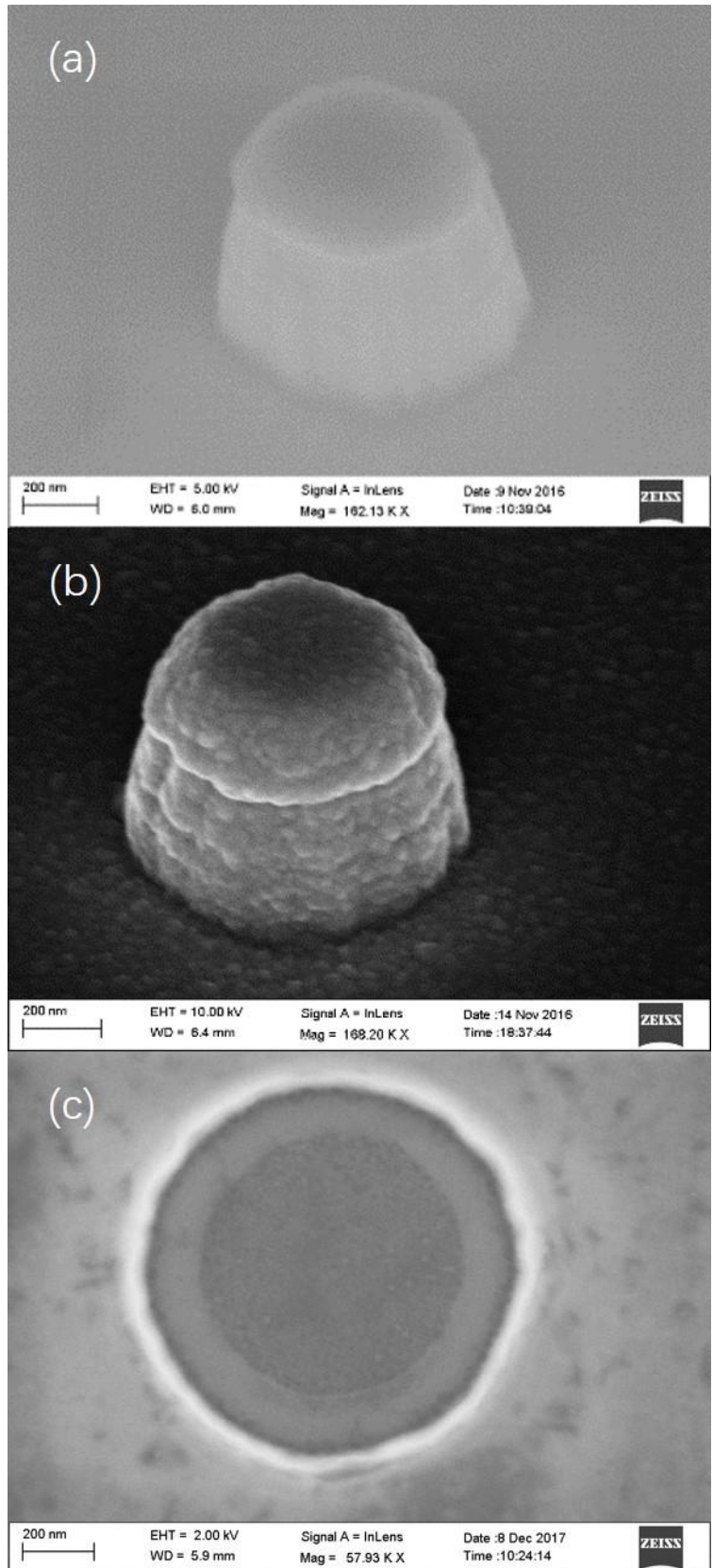

图 4-1 直径 900 nm 器件在 SEM 照片:(a) 干法刻蚀并去除残留掩模后，(b) 磁控溅射沉积 Ag 薄膜后，(c)器件制备完成后

## 4.3 InGaAs 体材料纳米激光器室温下光学性能表征

对基于 InGaAs 体材料的半导体-金属纳米激光器的光学性能测试同样在实验室自行搭建的 PL 测试系统进行，针对不同尺寸的器件，分别在室温和 77 K 的低温条件下进行了测试。我们在室温下对直径 900 nm 的器件进行了激光特性的测试，发现此规格的器件有两个波长的激光模式。一个模式的波长是在 1560 nm 附近，另一个在 1460 nm 附近。两个模式的光学场分布如图 4-2 所示的数值仿真结果所示。由于器件侧壁不陡直，这两个模式并不是典型的 SPP 模式或介质模式，而是每个模式都有 SPP 模式的成分和介质模式的成分。其中 1560 nm 的模式 SPP 成分占主导地位，而 1460 nm 的模式中介质成分占主导地位。仿真计算结果显示 1560 nm 模式的品质因子为 726，1460 nm 模式的品质因子为 383。在我们测试的 50 个器件中，46 个器件产生了 1560 nm 的激光，2 个器件产生了 1460 nm 的激光，另有 2 个器件能够同时产生两个波长的激光。我们将顺序呈现此三种器件的激光性能测试结果并进行分析。

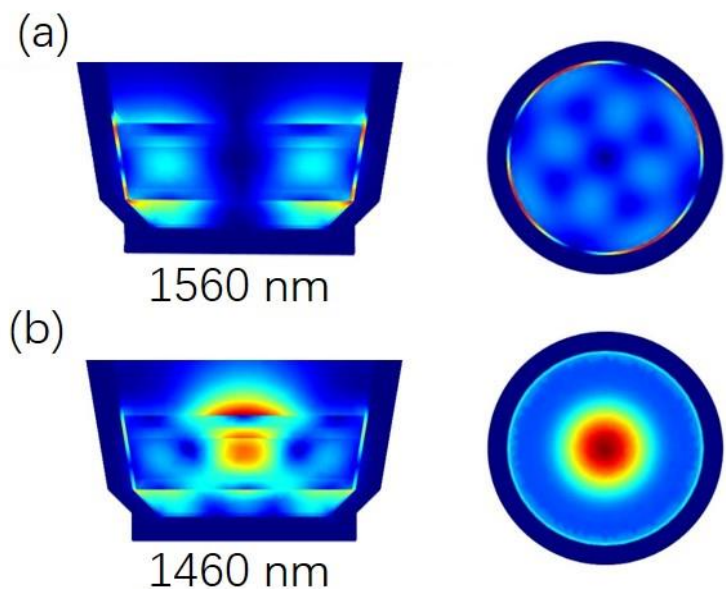

图 4-2 直径 900 nm 器件(a)1560 nm，(b)1460 nm 光学模式仿真

图 4-3 是一个产生 1560 nm 激光的器件的归一化 PL 光谱随泵浦光功率提高的演化过程。可以观察到，随泵浦功率提高，在波长 1560 nm 处出现光学模式峰，峰强度迅速增高，线宽也随之变窄。表现出明显的激光特性。

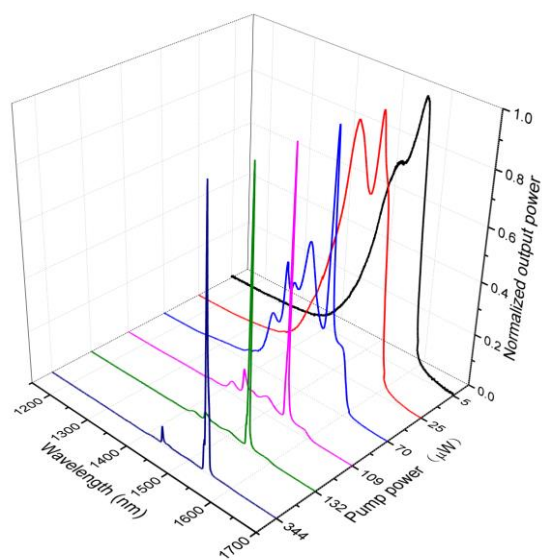

图 4-3 直径 900 nm 器件的 1560 nm 激光的光谱演化

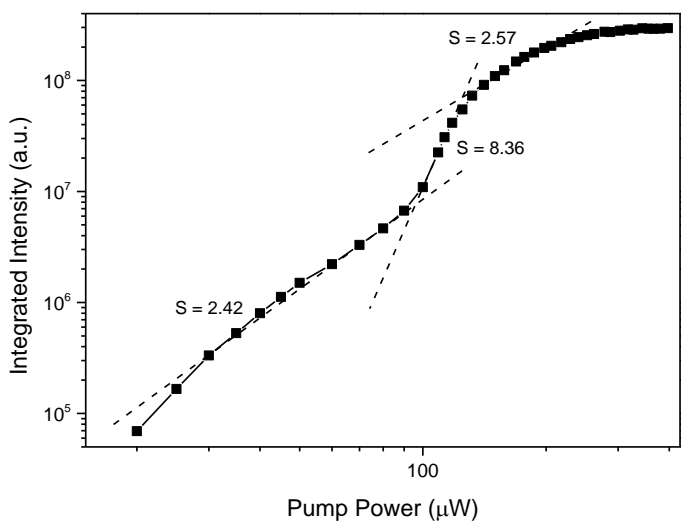

图 4-4 直径 900 nm 器件的 1560 nm 激光的输入输出曲线

图 4-4 是 1560 nm 激光的输入输出曲线，在双对数坐标系呈明显的 S 型，表现出一个激射过程。但是该曲线在自发辐射，自发辐射放大和受激辐射阶段的斜率分别为 2.42，8.36 和 2.57，其中自发辐射和受激辐射阶段的斜率表明这两个阶段模式的输出功率和泵浦强度呈接近平方的对应关系。这不是一个典型的对应于辐射复合的关系曲线，在双对数坐标系中约为 2 的斜率预示着非辐射复合成为发光的主要来源为非辐射复合。在图 4-5 所示的包含背景 PL 的输入输出曲线中，自发辐射阶段的斜率为

1.85，更进一步确定了非辐射复合的存在。我们认为非辐射复合是来源于微纳加工过程产生了材料的缺陷态和表面态的发光。这一猜想得到了实验的证实。图 4-6 为双对数坐标系中表示的未经加工的体材料薄膜，完成干法刻蚀后的纳米结构和完成 $Al_2O_3$ 沉积后的纳米结构的 PL 发光的输入输出曲线。这些曲线所表示的均为自发辐射的过程，可以发现对应于未经加工的薄膜材料的曲线的斜率为 1.07，表明 PL 输出功率与泵浦功率为线性关系，为典型的辐射复合的特征。而两种纳米结构的曲线的斜率均在 1.5 左右，表明辐射复合与非辐射复合同时存在，证实了微纳加工过程造成非辐射复合产生的猜想。

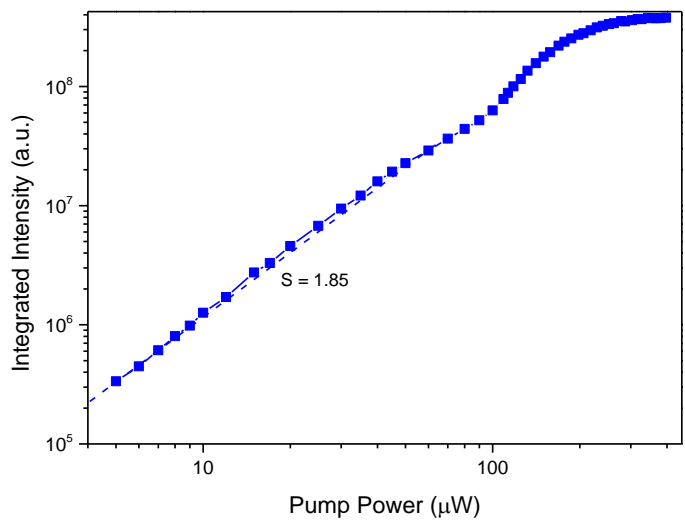

图 4-5 产生 1560 nm 激光的直径 900 nm 器件 PL 的输入输出曲线

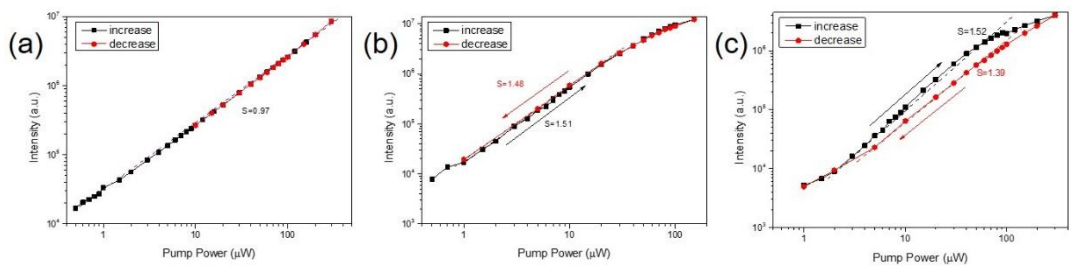

图 4-6 (a)体材料薄膜，(b)完成干法刻蚀后的纳米结构和(c)完成 $Al_2O_3$ 沉积后的纳米结构的 PL 发光的输入输出曲线

图 4-7 是 1560 nm 激光模式线宽与泵浦功率的关系曲线。可以观察到模式线宽是随着泵浦功率提高而减小的，直到焦耳热效应导致激光线宽开始展宽为止。从图 4-8 所示的光学模式发光峰的发光强度的倒数与

线宽的关系曲线,可以观察两者的关系同样符合激光的 Schawlow-Townes 方程。

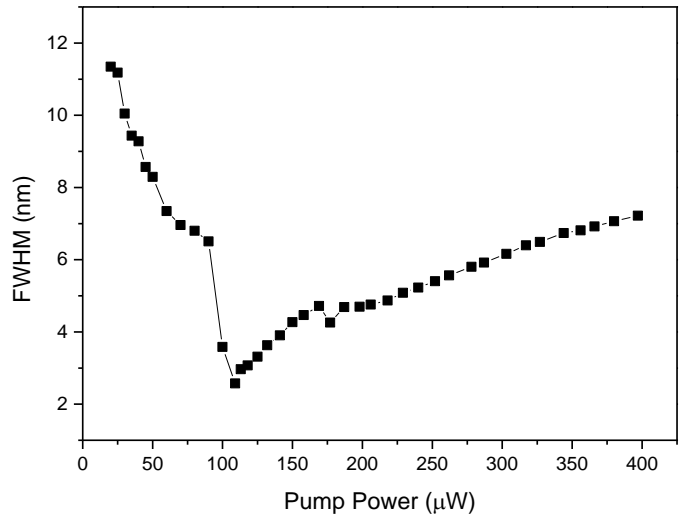

图 4-7 直径 900 nm 器件的 1560 nm 激光线宽与泵浦功率的关系曲线

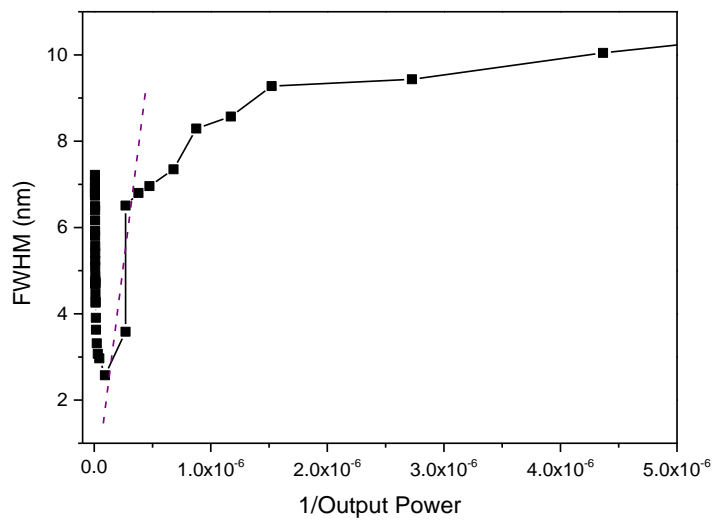

图 4-8 直径 900 nm 器件 1560 nm 模式发光强度倒数与线宽的关系曲线

图 4-9 是一个产生 1460 nm 激光的器件的归一化 PL 光谱随泵浦光功率提高的演化过程。可以观察到,随泵浦功率提高,在波长 1560 nm 处先出现光学模式峰,但是这个模式却没有发生激射。在泵浦功率达到 60 μW 左右,在 1460 nm 处出现模式峰并且强度迅速增高,线宽也随之变

窄，发生激射。

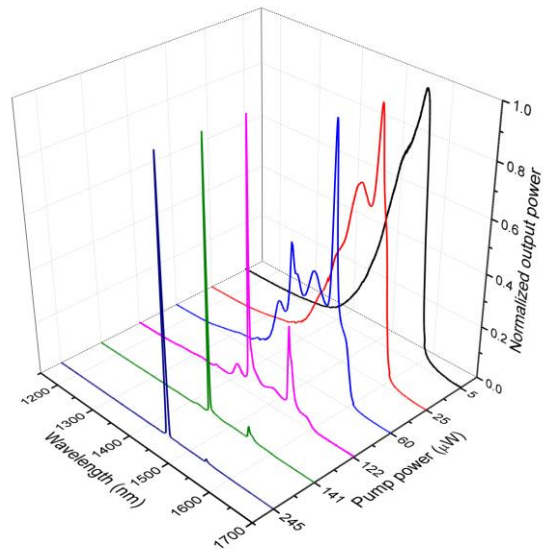

图 4-9 直径 900 nm 器件的 1460 nm 激光的光谱演化

图 4-10 是 1460 nm 激光的输入输出曲线，在双对数坐标系呈明显的 S 型，表现出一个激射过程。该曲线在自发辐射，自发辐射放大和受激辐射阶段的斜率分别为 4.09，16.70 和 2.57，其中自发辐射阶段的斜率远大于 2，达到如此高的斜率，应该源于除辐射种类外的其他原因。在图 4-11 所示的包含背景 PL 的输入输出曲线中，自发辐射阶段的斜率为 1.77，回归到小于 2 的范围，更证实了高斜率的产生有 1460 nm 激光模式自身的原因。为了探索这一原因，我们对未经加工的体材料薄膜，完成干法刻蚀后的纳米结构和完成 $Al_2O_3$ 沉积后的纳米结构的 PL 光谱的变化，其中经过微纳加工的两种结构与前文测试的激光器有同样的加工参数。图 4-12 是这三种结构的 PL 谱随泵浦光功率提高的演化过程，可以观察到未经加工的薄膜的 PL 光谱只表现出发光强度的提高，光谱的形状没有改变，表明对于每个波长的光子在这一测试的过程中产生的概率没有发生变化，材料的发光机制没有发生变化。而经过微纳加工的半导体纳米圆盘结构的 PL 光谱随着泵浦功率提高不仅有发光强度提高，光谱形状也发生了变化。当泵浦功率达到 50 μW 后，在 1550 nm 附近的发光强度增长速度开始减慢，在泵浦功率达到 200 μW 后甚至出现饱和的趋势。同时，在 1450 nm 处的发光强度开始快速的增强，直到强度超过 1550 nm 波长。这一变化可以理解为纳米级的材料尺寸导致 1550 nm 波长对应的能级在高泵浦下被填充饱和，而对应于 1450 nm 的一个能级开始被填充，带来了 1450 nm 附近的快速增长的增益，支持了 1460 nm 模式在自发辐射阶段的极快

速增长。

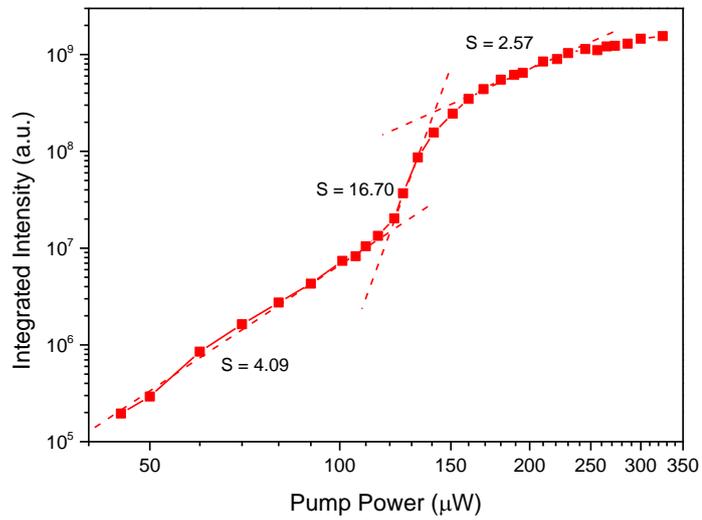

图 4-10 直径 900 nm 器件的 1460 nm 激光的输入输出曲线

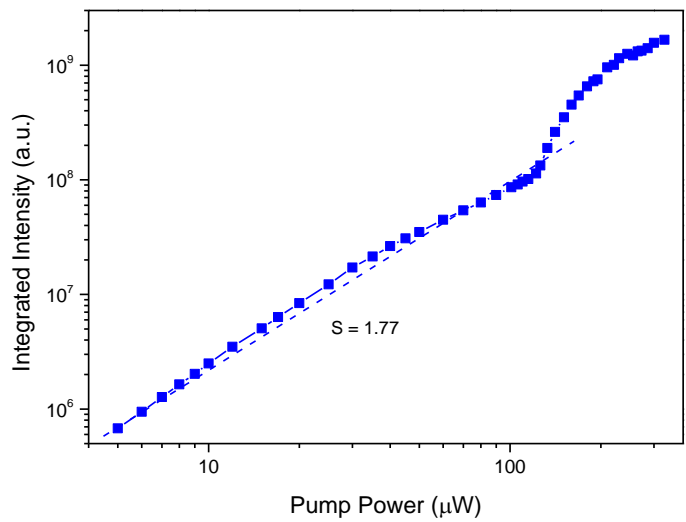

图 4-11 产生 1460 nm 激光的直径 900 nm 器件 PL 的输入输出曲线

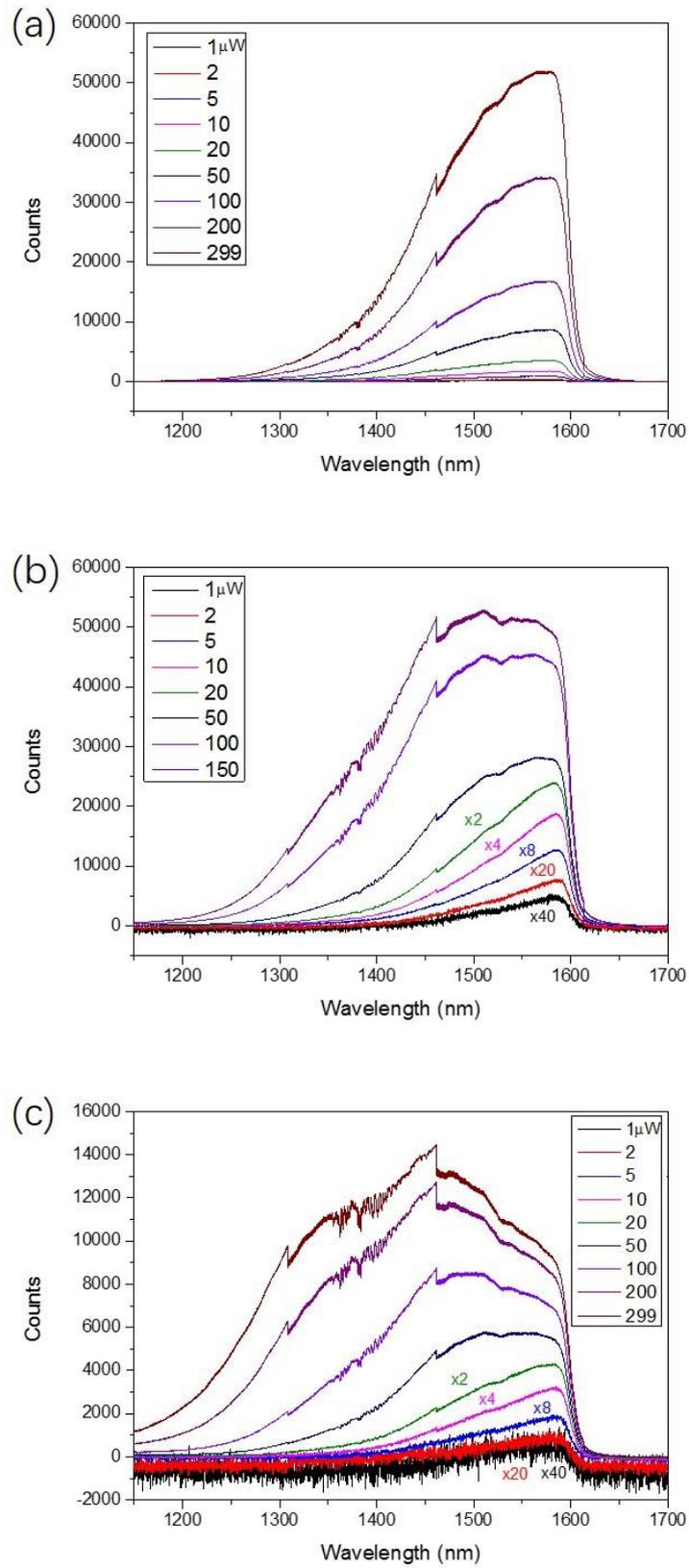

图 4-12 (a)体材料薄膜，(b)完成干法刻蚀后的纳米结构和(c)完成 Al$_2$O$_3$ 沉积后的纳米结构的 PL 光谱演化

图 4-13 是 1460 nm 激光模式线宽与泵浦功率的关系曲线。与 1560 nm 的激射过程相同，可以观察到模式线宽是随着泵浦功率提高而减小的，直到焦耳热效应导致激光线宽开始展宽为止。从图 4-14 所示的光学模式发光峰的发光强度的倒数与线宽的关系曲线，可以观察两者的关系同样符合激光的 Schawlow-Townes 方程。

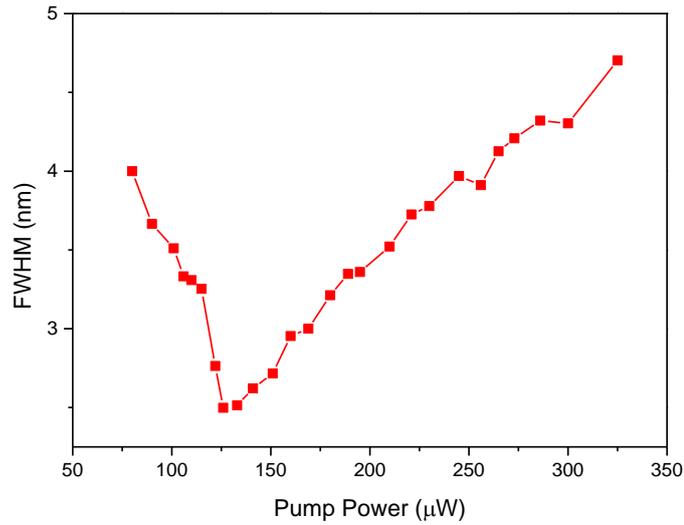

图 4-13 直径 900 nm 器件的 1460 nm 激光线宽与泵浦功率的关系曲线

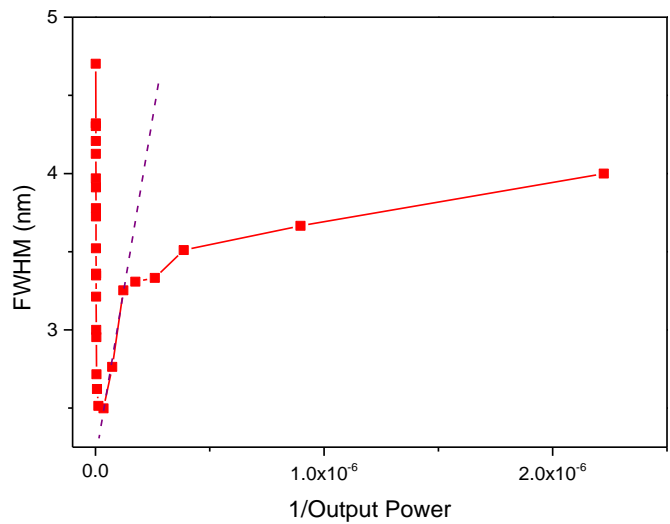

图 4-14 直径 900 nm 器件 1460 nm 模式发光强度倒数与线宽的关系曲线

对于可以同时支持 1560 nm 和 1460 nm 两个模式激光的器件会表现出两个模式的竞争。图 4-15 是同时得到两个激光模式器件的归一化 PL

光谱随泵浦功率提高的演化过程。从图中可以看到，随着泵浦功率提高，1560 nm 的光学模式首先出现并逐渐增强。在泵浦功率达到 80 μW 后，1460 nm 的光学模式出现，两个模式一起增强。随泵浦功率继续提高，1560 nm 的模式先达到激光阈值开始激射。之后 1460 nm 的模式达到阈值并激射最终由于 1460 nm 的激光会得到更多的增益而产生高于 1560 nm 激光的强度。

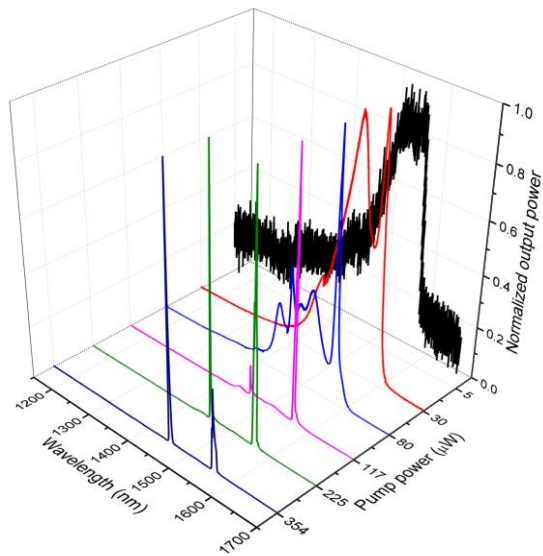

图 4-15 直径 900 nm 器件的 1560 nm 和 1460 nm 激光的光谱演化

图 4-16 为双对数坐标系中 1560 nm 和 1460 nm 激光的输入输出曲线。1560 nm 激光的曲线和只产生 1560 nm 激光的器件的输入输出曲线表现出相同的变化过程和特点。而 1460 nm 激光的输入输出曲线，则与只有 1460 nm 激光的器件有所不同，主要体现在自发辐射放大阶段的斜率没有明显增大，在部分阶段甚至略有下降。这可以理解为因为两个模式的阈值功率接近，而从能级考虑，低泵浦下 1560 nm 的模式相对容易得到更多的增益，1460 nm 的模式得到的增益较少，所以对增益的竞争导致了 1460 nm 模式的自发辐射放大被抑制。在高泵浦下，1550 nm 波长对应的能带饱和，1560 nm 激光得到的增益也达到饱和，1560 nm 的激光强度不再提高。同时，1450 nm 波长的能带的填充使得 1460 nm 激光得到更多的增益，最终其输出强度在泵浦功率达到 300 μW 后超过 1560 nm 的激光。图 4-17 的包含背景 PL 的输入输出曲线在 300 μW 以上的泵浦功率下并没有表现出输出功率饱和的现象，表明 1560 nm 激光在 300 μW 以上输出功率的饱和的原因是增益的饱和而不是器件的损坏。

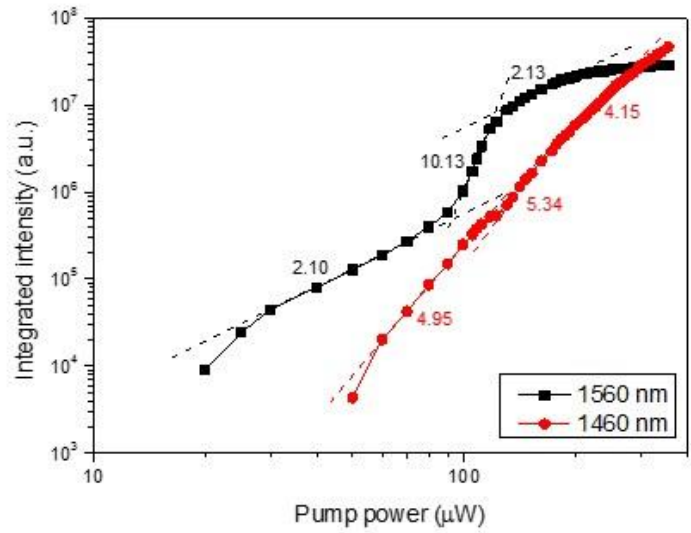

图 4-16 直径 900 nm 器件的 1560 nm 和 1460 nm 激光的输入输出曲线

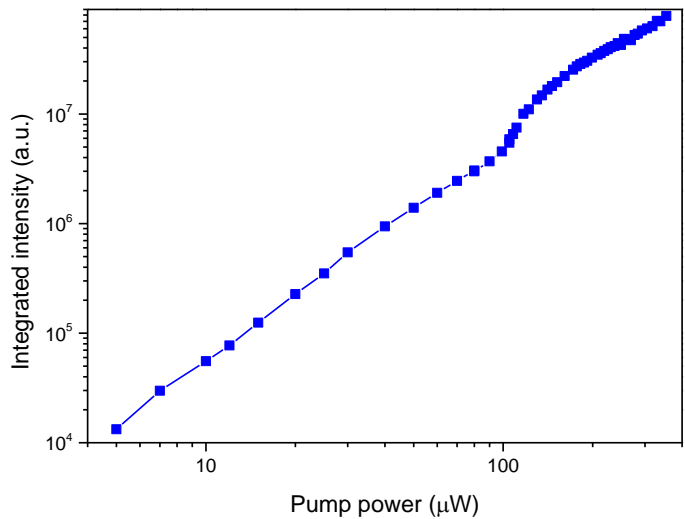

图 4-17 产生双波长激光的直径 900 nm 器件 PL 的输入输出曲线

图 4-18 和图 4-19 的 1560 nm 和 1460 nm 激光模式线宽与泵浦功率的关系曲线及发光强度的倒数与线宽的关系曲线，可以观察两者的关系同样符合激光的线宽变化特点和 Schawlow-Townes 方程。

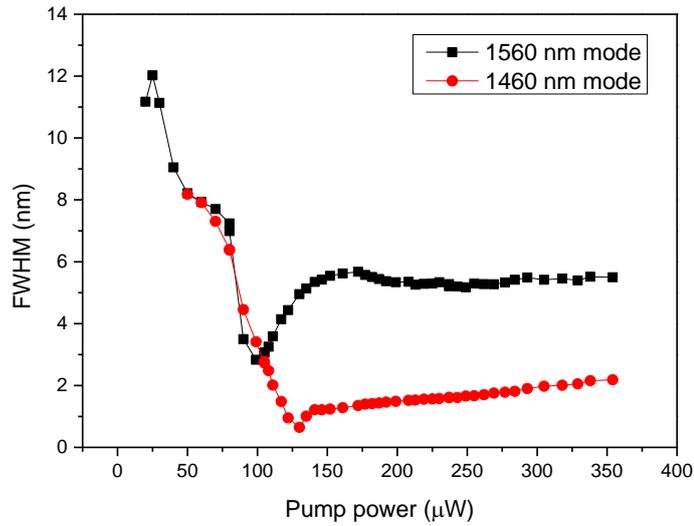

图 4-18 直径 900 nm 器件的 1560 nm 和 1460 nm 激光线宽与泵浦功率的关系曲线

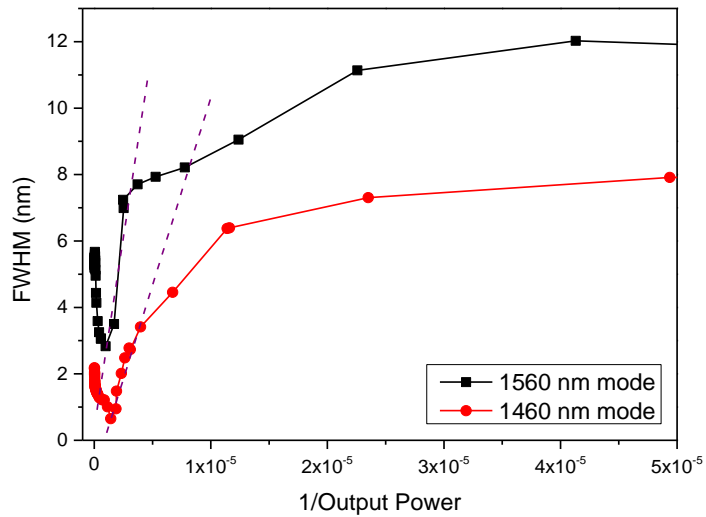

图 4-19 直径 900 nm 器件 1560 nm 和 1460 nm 模式发光强度倒数与线宽的关系曲线

至此，我们对该类型纳米激光器的激光性能及模式竞争可以得到如下分析结论。对于多数 900 nm 直径的器件而言，1560 nm 的激光具有更低的阈值增益，并且在低泵浦下由于波长接近增益波长而得到更多的增益。同时 1450 nm 附近的增益在器件损坏前尚未达到 1460 nm 模式的所需的极高的阈值增益，器件就只产生 1560 nm 的激光。少数器件 1560 nm

的模式由于某种原因被抑制，于是 1460 nm 的模式获得更多的增益产生激射。另有少数器件，两个模式具有接近的阈值增益，因此在 1560 nm 模式激射后，能带填充带来的 1450 nm 的增益会使 1460 nm 的模式激射。

通过对基于 InGaAs 体材料的半导体-金属复合圆盘腔纳米激光器的室温下光学性能表征，我们发现即使是体材料作为增益材料，在体积的限制下，依然存在高泵浦下低能级被填充饱和，高能级继续被填充的现象。这一现象造成了增益向短波长移动，支持了短波长的激射。对于这一现象，我们应该考虑到，当增益材料的体积继续减小，低能级对应的增益存在即使在器件被高泵浦损坏，也不能达到激射的阈值增益的可能。因此，在亚微米的尺寸下，增益材料的效率也开始成为半导体红外纳米激光器微型化的限制条件之一。

### 4.4 InGaAs 体材料纳米激光器低温下光学性能表征

具有更小尺寸的器件由于电子束曝光面积小，时间短，从而有效避免了电子束在电子束抗蚀剂中散射，使器件侧壁保持了良好的陡直度，如图 4-20 所示的圆盘直径 400 nm 的器件的 SEM 照片。这一尺寸的器件的有源区体积很小，在室温泵浦下，提供的增益尚未达到模式的阈值增益时，器件已由于焦耳热效应而损坏。因此，直径 400 nm 的器件的光学性能测试在 77 K 的温度下进行。

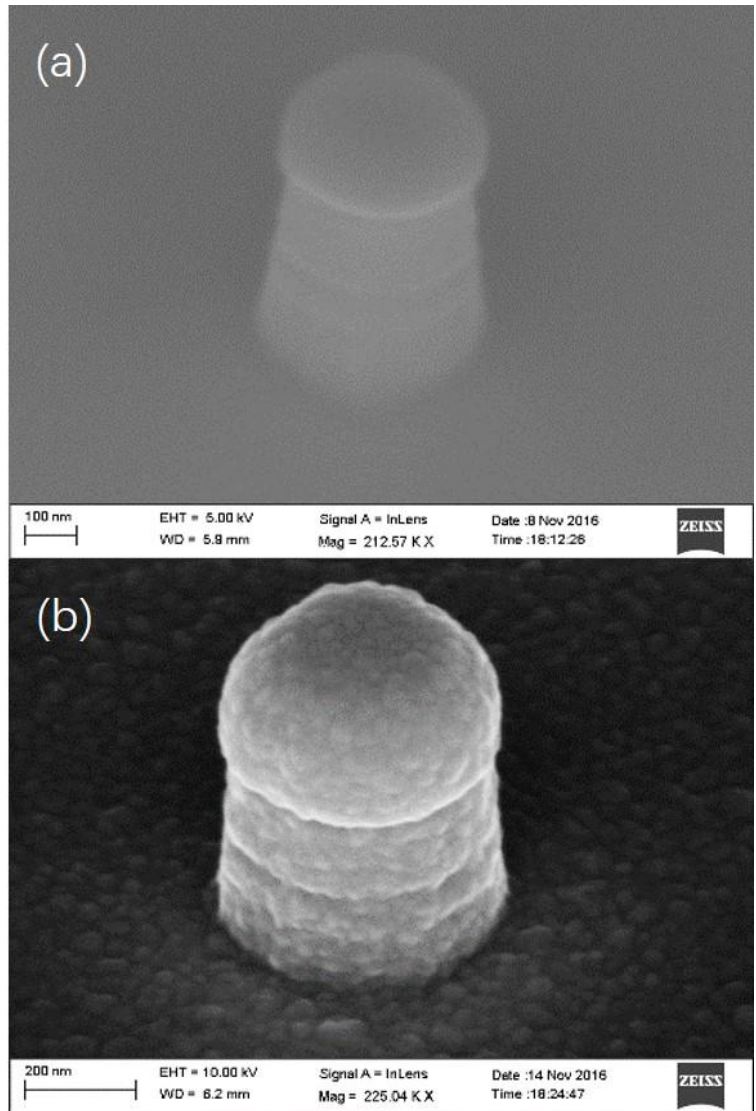

图 4-20 直径 400 nm 器件在 SEM 照片:(a) 干法刻蚀并去除残留掩模后,(b) 磁控溅射沉积 Ag 薄膜后

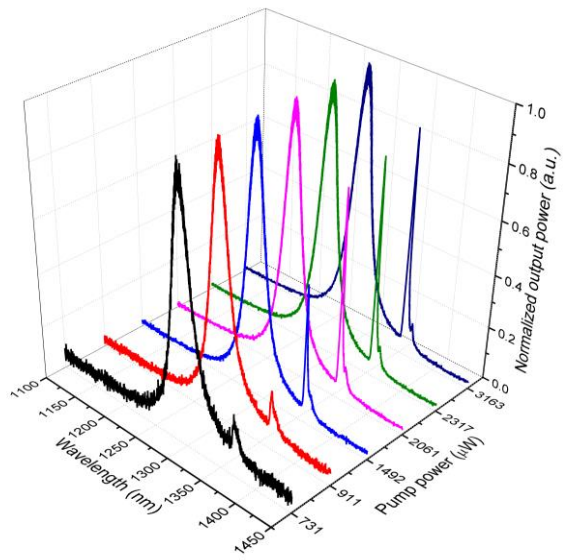

图 4-21 直径 400 nm 器件的 1360 nm 激光的光谱演化

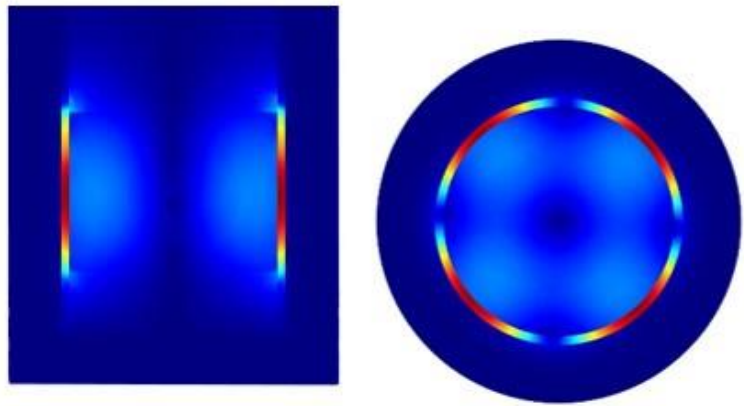

图 4-22 直径 400 nm 器件 1360 nm 光学模式仿真

图 4-21 是一个 400 nm 直径的器件的归一化 PL 光谱随泵浦光功率提高的演化过程。可以看到当泵浦功率达到 500 μW 以上，在 1360 nm 附近出现一个光学模式，随着泵浦功率提高，模式的强度不断增强，产生激光。通过数值仿真，我们认为该器件该波长的模式为如图 4-22 所示的 $TE_{21}$ SPP 模式，属于类 TE 模表面等离极化激元模式。

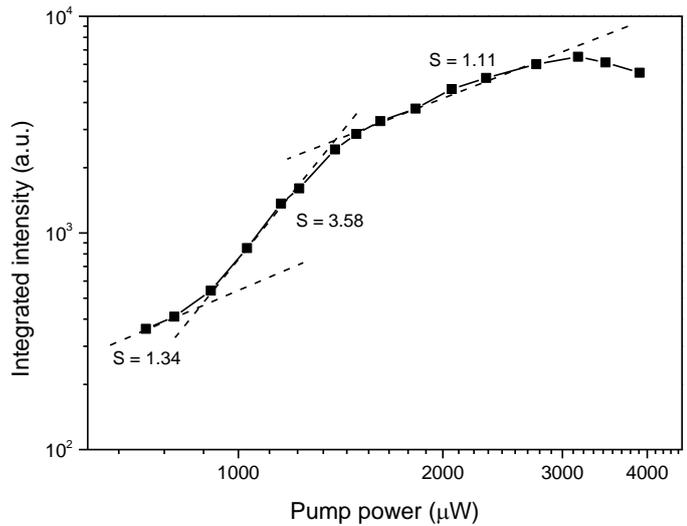

图 4-23 直径 400 nm 器件的 1360 nm 激光的输入输出曲线

图 4-23 是双对数坐标系中激光的输入输出曲线，在双对数坐标系呈明显的 S 型，表现出一个激射过程。该曲线在自发辐射，自发辐射放大和受激辐射阶段的斜率分别为 1.34，3.58 和 1.11，为典型的辐射复合的激射过程，表明在低温下缺陷发光受到了抑制。

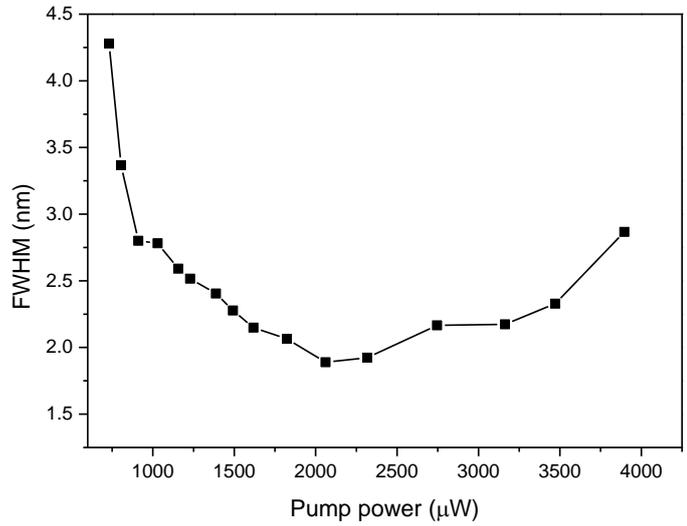

图 4-24 直径 400 nm 器件的 1360 nm 激光线宽与泵浦功率的关系曲线

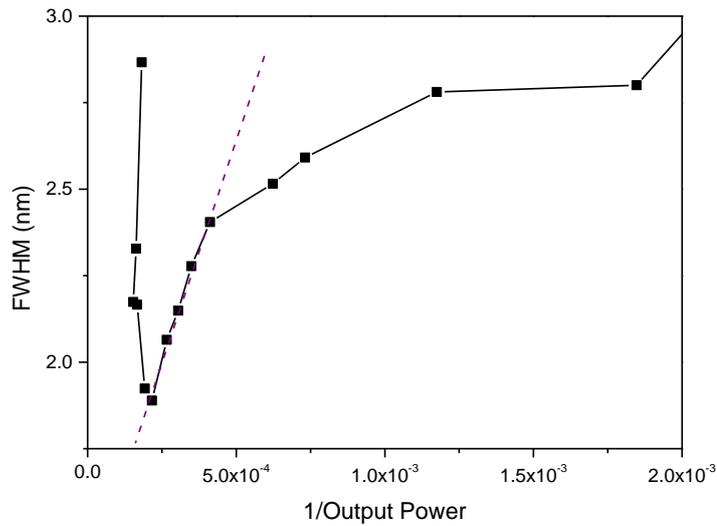

图 4-25 直径 400 nm 器件 1360 nm 模式发光强度倒数与线宽的关系曲线

图 4-24 和图 4-25 为激光模式线宽与泵浦功率的关系曲线及发光强度的倒数与线宽的关系曲线，可以观察两者的关系同样符合激光的线宽变化特点和 Schawlow-Townes 方程。

**4.5 本章小结**

本章介绍了以 InGaAs 体材料为增益材料的半导体-金属复合圆盘腔纳米激光器的实验表征。器件的形貌表征显示半导体圆盘直径 900 nm 的器件形貌较好，但是因为电子束曝光存在的电子散射，圆盘腔存在侧壁不陡直的问题。而对于直径 400 nm 的器件，电子散射不严重，不存在侧壁陡直度问题。在室温下的光学性能表征中直径 900 nm 的器件可以产生 1560 nm 和 1460 nm 两个波长的激光。我们发现对于这一类器件，在高泵浦功率下，由于小体积的限制效应，存在与多量子阱器件同样的低能级填充饱和，高能级被填充的现象，而 1460 nm 的激射的增益便来自这一填充过程。而在 77 K 的温度下的光学表征中，直径 400 nm 的器件产生了类 TE 模表面等离极化激元模式的激光。

# 第五章 结 论

本研究设计、制备了基于 InGaAsP 多量子阱和 InGaAs 体材料薄膜的半导体-金属复合圆盘腔纳米激光器，并且对激光器的激光性能进行了表征。研究结论如下：

1. 在半导体-金属复合腔纳米激光器的设计中可通过在增益介质与金属壳层之间加入一薄层透明介质层有效减低金属损耗，提高表面等离极化激元模式的品质因子。这一类型的器件可以通过标准化的微纳加工工艺批量制备。

2. 以 InGaAsP 多量子阱为增益材料的半导体-金属复合圆盘腔纳米激光器在室温下可以产生表面等离极化激元模式的激光。610 nm 直径的器件的激光模式为类 TM 模表面等离极化激元模式。因为该类器件体积极小，在高泵浦功率下，多量子阱势阱能级被填充饱和，载流子会填充势阱的子带能级和势垒能级，产生新的增益和支持对应波长的激射。对于圆盘直径 1 μm 的器件，可以先后产生波长 1550 nm、1310 nm 和 1180 nm 的激光。

3. 以 InGaAs 体材料为增益材料的亚微米级的半导体-金属复合圆盘腔纳米激光器在室温下同样可以激射。直径 900 nm 的器件可以产生 1560 nm 和 1460 nm 两个波长的激光。由于体积限制效应，该类器件存在与多量子阱器件同样的低能级填充饱和，高能级被填充的现象。而在 77 K 的低温光学表征中，直径 400 nm 的器件可以产生类 TE 模表面等离极化激元模式的激光。

# 致 谢

本研究工作是在甯存政教授的指导下完成。甯老师在研究中对实验结果中细节的深入思考、敏锐洞察让我受益良多。他深厚的理论功底，严谨的治学态度和对学术问题不懈思考与求索的学者风范令我敬仰。在他的指导下进行科学研究工作是我这两年来最荣幸的事情！

非常感谢课题组老师李永卓、孙皓、甘霖，博士后梁璋，博士章建行、冯家斌、王震、张聃旸、张琪瑶、钟子钊，工程师黄丹丹、郑熠泽对我的研究工作给予的帮助！在与他们每个人的交流与讨论中我学习到了许多的知识。

感谢清华大学电子系给予我博士后工作的机会！感谢中国科学院半导体研究所提供微纳加工工艺实验场地和设备！感谢中国科学院半导体研究所帮助我进行器件加工的工程师们及为我提供了实验材料的阚强、张瑞康老师！

感谢我的父母、亲人和朋友们一直以来对我的支持！

# 参考文献：

[1] 宁存政, 半导体纳米激光. 物理学进展, 2011, 31(3): 145~160.

[2] Ding K, Diaz J O, Bimberg D, et al, Modulation bandwidth and energy efficiency of metallic cavity semiconductor nanolasers with inclusion of noise effects. Laser & Photonics Reviews, 2015, 9(5): 488~497.

[3] Huang M, Mao S, Feick H, et al, Room-temperature ultraviolet nanowire nanolasers. Science, 2001, 292(5523): 1897~1899.

[4] Johnson C J, Yan H, Yang P, et al, Optical cavity effects in ZnO nanowire lasers and waveguides. J. Phys. Chem. B, 2003, 107(34): 8816~8828.

[5] Pauzauskie P J, Sirbuly D J, Yang P, Semiconductor nanowire ring resonator laser. Phys. Rev. Lett., 2006, 96(14): 143903.

[6] Agarwal R, Barrelet C J, Lieber C M, Lasing in single cadmium sulfide nanowire optical cavities. Nano Lett., 2005, 5(5): 917~920.

[7] Roder R, Wille M, Geburt S, et al, Continuous wave nanowire lasing. Nano Lett., 2013, 13(8): 3602~3606.

[8] Oulton P F, Sorger V J, Zentgraf T, et al, Plasmon laser at deep subwavelength scale. Nature, 2009, 461: 629~632.

[9] Liu X, Zhang Q, Yip J N, et al, Wavelength tunable single nanowire lasers based on surface plasmon polariton enhanced Burstein-Moss effect. Nano Lett., 2013, 13(11): 5336~5343.

[10] Levi A F J, McCall S L, Pearton S J, et al, Room temperature operation of submicrometre radius disk laser. Electron. Lett., 1993, 29(18): 1666~1667.

[11] Baba T, Fujita P, Sakai A, et al, Lasing characteristics of GaInAsP-InP strained quantum-well microdisk injection lasers with diameter of 2-10 μm. IEEE Photon. Technol. Lett., 1997, 9(7): 878~880.

[12] Srinivasan K, Borcelli M, Painter O, et al, Cavity Q, mode volume, and lasing threshold in small diameter AlGaAs microdisks with embedded quantum dots. Opt. Express, 2006, 14(3): 1094~1105.

[13] Park H, Kim S, Kwon S, et al, Electrically driven single-cell photonic crystal laser. Science, 2004, 305(5689): 1444~1447.

[14] Tomljenovic-Hanic S, de Sterke C M, Steel M J, et al, High-Q cavities in multilayer photonic crystal slabs. Opt. Express, 2007, 15(25): 17248~17253.

[15] Danner A J, Lee J C, Raftery J J, et al, Coupled-defect photonic crystal vertical cavity surface emitting lasers. Electron. Lett., 2003, 39(18): 1323~1324.

[16] Nozaki K, Watanabe H, Baba T, Photonic crystal nanolaser monolithically integrated with passive waveguide for effective light extraction. Appl. Phys. Lett., 2008, 92: 021108.

[17] Ding K, Ning C Z, Fabrication challenges of electrical injection metallic cavity semiconductor nanolasers. Semicond. Sci. Technol., 2013, 28: 124002.

[18] Hill M T, Oei Y S, Smalbrugge B, et al, Lasing in metallic-coated nanocavities. Nature Photon., 2007, 1(10): 589~594.

[19] Ding K, Hill M T, Liu Z C, et al, Record performance of electrical injection sub-wavelength metallic-cavity semiconductor lasers at room temperature. Opt. Express, 2013, 21(4): 4728~4733.

[20] Khajavikhan M, Simic A, Katz M, et al, Thresholdless nanoscale coaxial lasers. Nature, 2012, 482: 204-207.

[21] Yu K, Lakhani A, Wu M C, Subwavelength metal-optic semiconductor nanopatch lasers. Opt. Express, 2010, 18(9): 8790~8799.

[22] Kwon S, Kang J, Seassal C, et al Subwavelength plasmonic lasing from a semiconductor nanodisk with silver nanopan cavity. Nano Lett., 2010, 10(9): 3679~3683.

[23] Kwon S, Kang J, Kim S, et al, Surface plasmonic nanodisk/nanopan lasers. IEEE J. Quantum Elect., 2011, 47(10): 1346~1353.

[24] Hill M T, Marell M, Leong E S P, et al, Lasing in metal-insulator-metal sub-wavelength plasmonic waveguides. Opt. Express, 2009, 17(13): 11107~11112.

[25] Ding K, Liu Z, Yin L, et al, Electrical injection, continuous wave operation of subwavelength-metallic cavity lasers at 260 K. Appl. Phys. Lett., 2011, 98(23): 231108.

[26] Ding K, Liu Z C, Yin L J, et al, Room-temperature continuous wave lasing in deep-subwavelangth metallic cavites under electrical injection. Phys. Rev. B, 2012, 85(4): 041301.

[27] Ding K, Ning C Z, Metallic subwavelength-cavity semiconductor nanolaser. Light-sci. appl., 2012, 1(7): e20.

[28] Ding K, Yin L, Hill M T, et al, An electrical injection metallic cavity nanolaser with azimuthal polarization. Appl. Phys. Lett., 2013, 102(4): 041110.

# 个人简历

## 学习经历

2003.9-2007.7，学士，材料科学与工程，天津大学材料科学与工程学院。
2007.9-2009.7，硕士，材料学，天津大学材料科学与工程学院。
2009.9-2012.11，博士，纳米光子学，法国里昂中央理工大学校(Ecole Centrale de Lyon)。

## 工作经历

2013.8-2015.12，**博士后**，压电光电子学，国家纳米科学中心，中国科学院北京纳米能源与系统研究所。
2016.1-今，**博士后**，电子科学与技术，清华大学电子系。

# 博士生期间发表的学术论文

1. **Taiping Zhang**, Ali Belarouci, Ségolène Callard, Pedro Rojo-Romeo, Xavier Letartre, Pierre Viktorovitch. Plasmonic-photonic hybrid nanodevice. *International Journal of Nanoscience* 2012, Vol. 11, Issue (4) : 1240019.

2. Daniele Costantini, Leo Greusard, Adel Bousseksou, R. Rungsawang, **Taiping Zhang**, Ségolène Callard, Jean Decobert, Francois Lelarge, G. Duan, Yannick De Wilde, Raffaele Colombelli. In situ generation of surface plasmons polaritons using a near-infrared laser diode. *Nano Letters* 2012, Vol. 12, Issue (9) : 4693. (sci/IF:12.712)

3. Ali Belarouci, **Taiping Zhang**, Than Phong Vo, Ségolène Callard, Pedro Rojo-Romeo, Xavier Letartre, Pierre Viktorovitch. Plasmonic-photonic hybrid nanodevice: A new route toward 3D light harnessing. DOI: 10.1109/ICTON.2011.5970774.

4. Taha Benyattou, Ali Bellarouci, Xavier Letartre, Emmanuel Gerelli, **Taiping Zhang**, Pierre Viktorovitch. New tracks toward 3D light harnessing: high Q Slow Bloch mode engineering and coupling to 0D nanophotonic structures. Proceedings of SPIE - The International Society for Optical Engineering 7941.

5. Ali Bellarouci, Taha Benyattou, Xavier Letartre, Pedro Rojo-Romeo, **Taiping Zhang**, Pierre Viktorovitch. 3D light harnessing based on coupling engineering between 1D-2D photonic crystal membranes and 0D photonic structures. DOI: 10.1109/ICTON.2010.5548968.

6. Peikang Bai, Shengliang Hu, **Taiping Zhang**, Jing Sun, Shirui Cao. Effect of laser pulse parameters on the size and fluorescence of nanodiamonds formed upon pulsed-laser irradiation. *Materials Research Bulletin* 2010, Vol. 45 : 826. (sci/IF:2.446)

# 博士后期间发表的学术论文

1. Renrong Liang, **Taiping Zhang**, Jing Wang, Jun Xu. High-resolution light-emitting diode array based on an ordered ZnO nanowire/SiGe heterojunction. *IEEE Transactions on Nanotechnology* 2016, Vol. 15, Issue (3) : 539. (共同一作, sci/IF:2.485)

2. **Taiping Zhang**, Renrong Liang, Lin Dong, Caofeng Pan, Jing Wang, Jun Xu. Wavelength tunable infrared light emitting diode based on ordered ZnO nanowire/$Si_{1-x}Ge_x$ alloy heterojunction. *Nano Research* 2015, Vol. 8, Issue (8) : 2676. (sci/IF:7.354)

3. **Taiping Zhang**, Ségolène Callard, Cécile Jamois, Céline Chevalier, Di Feng, Ali Belarouci. Plasmonic-photonic crystal coulped nanolaser. *Nanotechnology* 2014, Vol. 25, Number 31 : 315201. (sci/IF:3.440)

4. Mengxiao Chen, Caofeng Pan, **Taiping Zhang**, Xiaoyi Li, Renrong Liang, Zhong Lin Wang. Tuning light emission of a pressure-sensitive silicon/ZnO nanowires heterostructure matrix through piezo-phototronic effects. *ACS Nano* 2016, Vol. 10, Issue (6) : 6074. (sci/IF:13.942)

5. Haitao Liu, Qilin Hua, Ruomeng Yu, Yuchao Yang, **Taiping Zhang**, Yingjiu Zhang, Caofeng Pan. A bamboo-like GaN microwire-based piezotronic memristor. *Advanced Functional Materials* 2016, Vol. 26, Issue (29) : 5307. (sci/IF:12.124)

6. Renrong Liang, Jing Wang, Jun Xu, **Taiping Zhang**. High resolution light emitting diode array based on ordered ZnO nanowire/SiGe heterojunction. IEEE NANO-2015 15th International Conference on Nanotechnology, DOI: 10.1109/NANO.2015.7388687.

7. Xiaoyi Li, Mengxiao Chen, Ruomeng Yu, **Taiping Zhang**, Dongsheng Song, Renrong Liang, Qinglin Zhang, Shaobo Chen, Lin Dong, Anlian Pan, Zhong Lin Wang, Jing Zhu, Caofeng Pan. Enhancing light emission of ZnO-nanofilm/Si-micropillar heterostructure arrays by piezo-phototronic effect. *Advanced Materials* 2015, Vol. 27, Issue (30) : 4447. (sci/IF:19.791)

8. Chunfeng Wang, Rongrong Bao, Kun Zhao, **Taiping Zhang**, Lin Dong,

Caofeng Pan. Enhanced emission intensity of vertical aligned flexible ZnO nanowire/p-polymer hybridized LED array by piezo-phototronic effect. ***Nano Energy*** 2015, Vol. 14, : 364. (sci/IF:12.343)

9. Delphine Manchon, Jean Lermé, **Taiping Zhang**, Alexis Mosset, Cécile Jamois, Christophe Bonnet, Jan-Michael Rye, Ali Belarouci, Michel Broyer, Michel Pellarin, Emmanuel Cottancin. Plasmonic coupling with most of the transition metals: a new family of broad band and near infrared nanoantennas. ***Nanoscale*** 2015, Vol. 7, Issue (3) : 1181. (sci/IF:7.367)